\documentclass[aps,superscriptaddress,twocolumn]{revtex4-1}

\pdfoutput=1

\usepackage{graphicx}
\usepackage{amsmath}
\usepackage{amssymb}
\usepackage{verbatim}
\usepackage{braket}
\usepackage{color}
\usepackage{hyperref}
\hypersetup{
	colorlinks=true,
	linkcolor=red,      
	urlcolor=blue,
	citecolor=blue
}
\begin{document}
%\linenumbers

\title{Topologically nontrivial multicritical points}
	
\author{Ranjith R Kumar}
\email{ranjith.btd6@gmail.com}
\affiliation{Department of Physics, Indian Institute of Technology Bombay, Powai, Mumbai 400076, India.}
\author{Pasquale Marra}
\email{pasquale.marra@keio.jp}
\affiliation{Department of Engineering and Applied Sciences, Sophia University, 7-1 Kioi-cho, Chiyoda-ku, Tokyo 102-8554, Japan.}
\affiliation{Department of Physics, and Research and Education Center for Natural Sciences, Keio University, 4-1-1 Hiyoshi, Yokohama, Kanagawa, 223-8521, Japan.}
\affiliation{Graduate School of Informatics, Nagoya University, Furo-cho, Chikusa-Ku, Nagoya, 464-8601, Japan.}	
	
\date{\today}

\begin{abstract}
	Recently, the intriguing interplay between topology and quantum criticality has been unveiled in one-dimensional topological chains with extended nearest-neighbor couplings. In these systems, topologically distinct critical phases emerge with localized edge modes despite the vanishing bulk gap. In this work, we study the topological multicritical points at which distinct gapped and critical phases intersect. Specifically, we consider a topological chain with coupling up to the third nearest neighbors, which shows stable localized edge modes at the multicritical points. These points possess only nontrivial gapped and critical phases around them and are also characterized by the quadratic dispersion around the gap-closing points. We characterize the topological multicritical points in terms of the topological invariant obtained from the zeros of the complex function associated with the Hamiltonian. Further, we analyze the nature of zeros in the vicinity of the multicritical points by calculating the discriminants of the associated polynomial. The discriminant uniquely identifies the topological multicritical points and distinguishes them from the trivial ones. Moreover, we identify the underlying physical mechanism in terms of kinetic inversion in higher-order terms.
We finally study the robustness of the zero-energy modes at the multicritical points at weak disorder strengths, and reveal the presence of a topologically nontrivial gapless Anderson-localized phase at strong disorder strengths.
\end{abstract}

\maketitle

\section{Introduction}

The interplay between topology and quantum criticality has been recent interest in topological states of matter \cite{verresen2018topology,verresen2019gapless,jones2019asymptotic,verresen2020topology,rahul2021majorana,niu2021emergent,PhysRevB.104.075132,PhysRevResearch.3.043048,fraxanet2021topological,keselman2015gapless,scaffidi2017gapless,duque2021topological,kumar2021multi,kwwangSciPostPhys.12.4.134,kumar2021topological,kumar2023topological,PhysRevLett.129.210601,PhysRevA.109.062226,PhysRevB.110.045119,yu2024universal,PhysRevA.110.022212,zhou2025gapless,xue2025gapless,xue2025exploring,zhou2025topological,yang2025deconfined,kumar2025topological,cv5q-8t25}. Topology of electronic band structure dictates the distinct gapped phases in various topological models \cite{haldane1988model,kane2005quantum,wang2017topological}. These gapped phases are characterized in terms of topological invariants that are quantized and protected by the bulk gap \cite{thouless1982quantized}. In one-dimensional (1D) topological chains, the topological invariant derived from the bulk Bloch states remains quantized and can efficiently count the number of edge modes that are localized at each end of the open chain~\cite{hatsugai_chern_1993,ryu_topological_2002,teo_topological_2010}. This is the well-known bulk-boundary correspondence in topological states of matter \cite{hasan2010colloquium}. Upon tuning the system parameters, the gapped phases undergo a transition into another distinct gapped phase by changing the topological invariant via a topological phase transition at which the bulk gap closes. The localization lengths of edge modes of the gapped phases diverge at a transition or quantum critical point \cite{continentino2020finite}. Therefore, conventional bulk-boundary correspondence forbids the presence of localized edge modes at a quantum critical point.

However, if the model hosts topological phases with higher topological invariant number, then the quantum critical points separating the two nontrivial phases can host edge modes \cite{rahul2021majorana,verresen2020topology,kumar2021topological,kumar2023topological}. This interplay between topology and quantum criticality violates the conventional bulk-boundary correspondence \cite{verresen2018topology}. The conventional topological invariant, defined as the winding number of the bulk Bloch states, fails to capture the number of edge modes at a quantum critical phase. Therefore, an alternative definition of topological invariant based on the number of zeros and poles of a complex function associated with the Hamiltonian is used in Ref.~\cite{verresen2018topology}. This characterizes the topological properties of both gapped and quantum critical phases in terms of the zeros that fall inside the unit circle on a complex plane. Additionally, the number of zeros on the unit circle measures the properties of various quantum critical phases and identifies the corresponding conformal field theory (CFT). Therefore, complete characterization of both the topological and critical properties of various quantum phases of the model can be done using this method. Alternatively, topological quantum criticalities have been studied, e.g., using symmetry properties \cite{verresen2019gapless}, conformal boundary conditions \cite{PhysRevLett.129.210601}, correlations of string operators \cite{jones2019asymptotic}, and universal entanglement spectrum \cite{yu2024universal}.

Distinct quantum critical phases intersect at a special point known as multicritical points \cite{rufo2019multicritical,kumar2021multi,kumar2021topological,kumar2023topological}. These are the points at which two or more gapped or quantum critical phases meet in the parameter space. The multicritical points, in general, can be characterized either based on the number of gap-closing points in the Brillouin zone, since they inherit the critical properties of both intersecting phases, or based on the unique nature of dispersion, such as being quadratic near the gap-closing points. A topological phase transition between distinct quantum critical phases occurs at the multicritical points of both kinds, and the number of edge modes changes across these points. Therefore, the localization length of edge modes diverges as the parameters are tuned towards a multicritical point. This observation leads to the conclusion that the edge modes delocalize into the bulk at a multicritical point, and they remain topologically trivial.

In this work, we revisit such instances and revise the understanding by identifying topologically nontrivial multicritical points under certain constraints on the parameters. 
We intend to show that in a regime where the critical lines lie on a plane of the parameter space that only hosts topologically nontrivial phases 
(where all the gapped and critical phases are nontrivial), localized edge modes also occur at the multicritical points.
In other words, topologically trivial multicritical points correspond to the intersection of critical lines separating trivial and nontrivial gapped phases, with the edge modes that are not robust against disorder, since any perturbation can move the system away from the critical point into the trivial region. 
In contrast, in our work, we consider multicritical points that separate distinct gapless (critical) lines separating topological phases that are all nontrivial.
In this case, the edge modes at the multicritical points are topologically robust against disorder, since the system cannot be perturbed into a topologically trivial phase.
This is our definition of "topologically nontrivial multicritical points".
In other words, a multicritical point is "nontrivial" if it is not just a transition to a vacuum or a trivial state, but rather a junction where the system is forced to choose between several different nontrivial topological phases.
Interestingly, these multicritical points are characterized by quadratic bulk dispersion and therefore exhibit critical edge modes (or gapless phases) that go beyond the conventional Dirac-like critical behavior.
Moreover, the multicritical points are robust against weak disorder, while they vanish at stronger disorder strength, revealing the presence of a gapless and topologically nontrivial Anderson-localized phase characterized by the presence of two zero-energy modes localized at the end of the chain.
The unique features of multicritical points can be exploited to enhance topological quantum qubits\cite{10.21468/SciPostPhys.3.3.021}, where topologically protected localized edge modes carry the quantum information~\cite{nayak_non-abelian_2008,stern_topological_2013,das-sarma_majorana_2015,roy_topological_2017,lahtinen_a-short_2017,stanescu_introduction_2017,beenakker_search_2020,marra_Majorana_2022}. 

\section{Model and Topological Phase Diagram}

We consider spinless fermions in a one-dimensional (1D) lattice chain that represents topological insulators and superconductors, such as the Su–Schrieffer–Heeger model~\cite{su_solitons_1979} or the Kitaev  chain~\cite{kitaev_unpaired_2001}.
Crucially, we generalize these models considering longer-range couplings, i.e., hopping and superconducting pairing terms extended past the nearest neighbor~\cite{niu_majorana_2012,hsu2020topological,vodola_kitaev_2014,viyuela_topological_2016,alecce_extended_2017,kumar2021topological,kumar2023topological}. 
It is worth mentioning that, by considering systems with unbroken time-reversal symmetry (class BDI) and longer-range couplings~\cite{niu_majorana_2012,vodola_kitaev_2014,viyuela_topological_2016,alecce_extended_2017}, e.g., where both the hopping and superconducting pairing terms extend to the nearest and next nearest neighbors, one can realize topologically nontrivial phases with higher winding numbers $|w|>1$.
This corresponds to the existence of a number $|w|$ of topologically-protected end modes at the boundary, which are mutually orthogonal and physically distinguishable~\cite{niu_majorana_2012,alecce_extended_2017}.

\begin{figure}[t]
	\includegraphics[width=8.5cm,height=4.5cm]{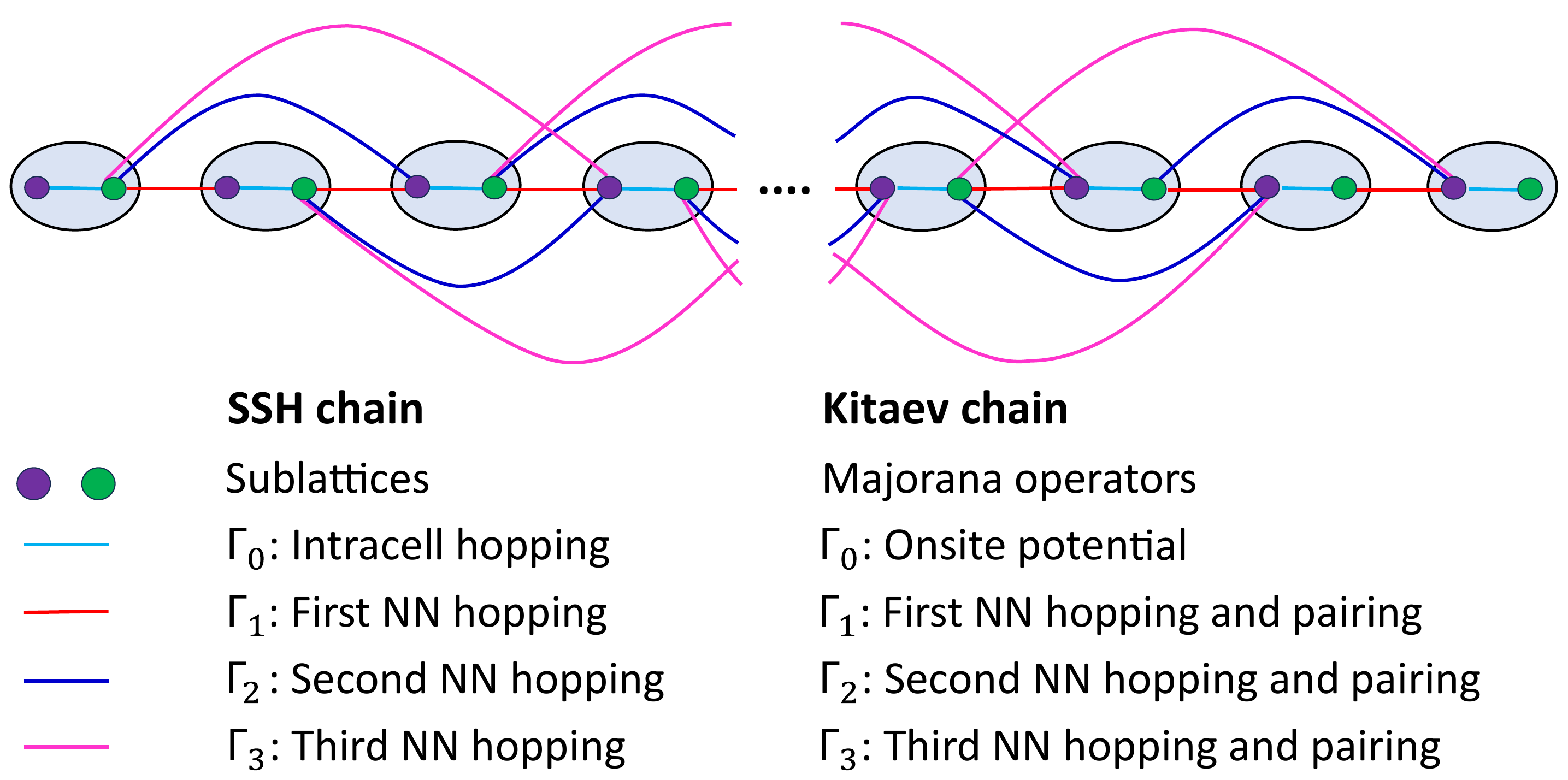}
	\caption{
	Schematic representation of the model under open boundary conditions with next-nearest coupling parameters for the extended Su–Schrieffer–Heeger (SSH) and Kitaev models.
	}
	\label{fig01}
\end{figure}

In momentum space, the two-band Bloch Hamiltonian can be written as
\begin{equation}
H(\boldsymbol{\Gamma},k)= \boldsymbol{\chi}(\boldsymbol{\Gamma},k)\cdot\boldsymbol{\sigma}
\end{equation}
where the parameter space $\boldsymbol{\Gamma}=\left\lbrace \Gamma_0,\Gamma_1,\Gamma_2,\Gamma_3 \right\rbrace$ and components of the Hamiltonian are $\chi_x(\boldsymbol{\Gamma},k)=\Gamma_0+\Gamma_1\cos(k)+\Gamma_2\cos(2k)+\Gamma_3\cos(3k)$, $\chi_y(\boldsymbol{\Gamma},k)=\Gamma_1\sin(k)+\Gamma_2\sin(2k)+\Gamma_3\sin(3k)$ and $\chi_z(\boldsymbol{\Gamma},k)=0$ with Pauli matrices $\boldsymbol{\sigma}=(\sigma_x,\sigma_y,\sigma_z)$. 
The parameters $\Gamma_{0,1,2,3}$ represent, respectively, intracell, intercell first, second, and third nearest neighbor hopping amplitudes in 1D Su-Schrieffer-Heeger (topological insulator) model \cite{su_solitons_1979}, with the tight-binding Hamiltonian
\begin{align}
H_\text{SSH} &= 
\Gamma_0 \sum_{i}  c_{i,a}^{\dagger} c_{i,b} + 
\Gamma_1 \sum_{i} (c_{i+1,a}^{\dagger} c_{i,b} + \text{h.c.}) \nonumber\\
&+ \Gamma_2 \sum_{i}( c_{i+2,a}^{\dagger} c_{i,b} + \text{h.c.})\nonumber\\
&+ \Gamma_3 \sum_{i}( c_{i+3,a}^{\dagger} c_{i,b} + \text{h.c.}) \label{SSH}
\end{align}
where $c^{\dagger}_{i,a(b)}$ and $c_{i,a(b)}$ are the fermionic creation and annihilation operators. The subscripts $a,b$
denote the sub-lattices. 
Similarly, the $\Gamma_{0,1,2,3}$ represents, respectively, the onsite potential, first, second. and third nearest neighbor hopping and pairing amplitudes in 1D Kitaev (topological superconductor) model \cite{kitaev_unpaired_2001}, with the tight-binding Hamiltonian
\begin{align}
H_\text{Kitaev} &= \Gamma_0 \sum_{i} ( 2 c_{i}^{\dagger}c_{i}-1) 
- \Gamma_1 \sum_{i } (c_{i}^{\dagger}c_{i+1} + c_{i}^{\dagger}c_{i+1}^{\dagger} + \text{h.c.}) \nonumber\\
&- \Gamma_2 \sum_{i } ( c_{i}^{\dagger}c_{i+2} +  c_{i}^{\dagger} c_{i+2}^{\dagger} + \text{h.c.}) \nonumber\\
&- \Gamma_3 \sum_{i} ( c_{i}^{\dagger}c_{i+3} +  c_{i}^{\dagger} c_{i+3}^{\dagger} + \text{h.c.})
\label{kitaev}
\end{align}
where the nearest neighbor hopping and pairing amplitudes are considered to be equal. A schematic representation of the model is shown in Fig.\ref{fig01} under open boundary condition.
These models, having hopping amplitudes extended up to the third nearest neighbors, can support topological phases with topological invariant up to three, corresponding to localized edge modes at each end of the open chain. 

\begin{figure}[t]
	\includegraphics[width=5.2cm,height=5cm]{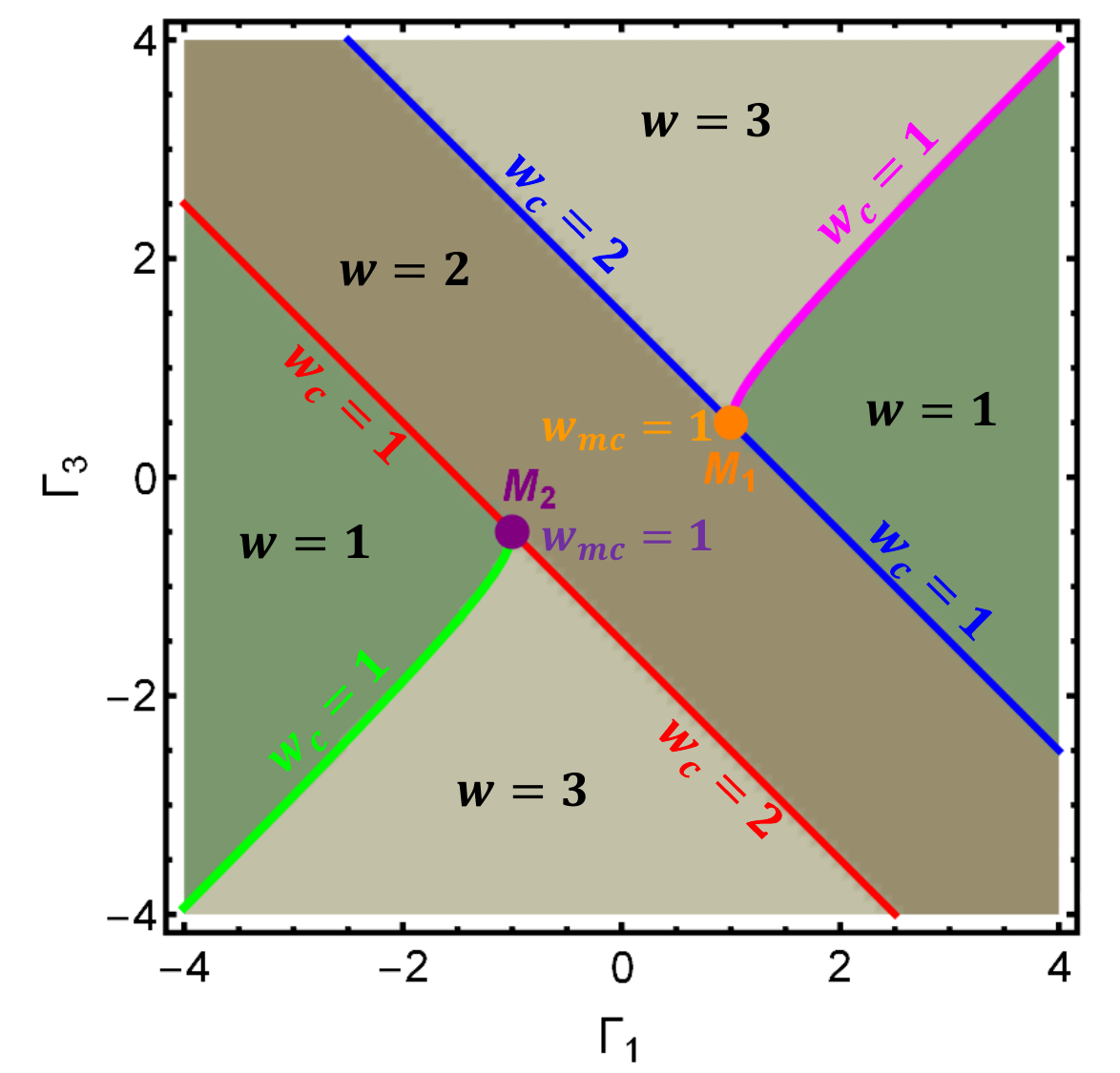}
	\caption{Topological phase diagram for $\Gamma_{0}=0.2$ and $\Gamma_{2}=1$. Topological gapped phases are identified with winding number $w=1,2,3$. These phases are separated by the critical lines colored red, blue, magenta and green. At these critical lines, the gap closing occurs at momenta $k_0=0$ for red, $k_0=\pm\pi$ for blue and at arbitrary $k_0$ points for magenta and green lines. The topological properties of the critical lines are identified with $w_c$. The red and blue critical lines possess two distinct phases with $w_c=1$ and $w_c=2$ (see Section~\ref{sec3}). The magenta and green lines are identified with $w_c=1$. At the intersection of the critical lines, the multicritical points are obtained and are named $M_{1,2}$. We identify non-trivial topological properties at these points given by $w_{mc}=1$. This is the main result of this work (see Section~\ref{sec4}).}
	\label{fig1}
\end{figure}

In general, topological properties such as edge states can be captured using the topological invariant number \cite{kane2005quantum,thouless1982quantized}. For 1D systems, winding of the Andersen pseudospin vector $\boldsymbol{\chi}(\boldsymbol{\Gamma},k)$ in the Brillouin zone gives an integer-valued winding number that efficiently counts the number of localized edge modes. The winding number can be obtained as \cite{hasan2010colloquium,chen2019universality}
\begin{equation}
w=\frac{1}{2\pi} \oint\limits_{BZ} F(\boldsymbol{\Gamma},k)= \frac{1}{2\pi} \oint\limits_{BZ} \frac{\chi_x\partial_k \chi_y - \chi_y \partial_k \chi_x}{\chi_x^2+\chi_y^2} dk
\label{windnum}
\end{equation}
where $w\in\mathbb{Z}$ at various gapped phases. The integrand is referred to as curvature function which can be written in terms of bulk Bloch bands as $F(\boldsymbol{\Gamma},k)= i\left\langle u_k|\partial_k|u_k\right\rangle $ with Bloch wave function $\psi_k(r)=u_k(r)e^{ikr}$. Therefore, according to the bulk-boundary correspondence, the value of $w$ physically quantifies the number of topologically protected edge modes in an open chain. A topological phase diagram can be obtained based on the values of $w$ as shown in Fig.~\ref{fig1}. The phase transition between distinct gapped phases involves bulk gap closing as
\begin{equation}
E(\boldsymbol{\Gamma},k)=\pm\sqrt{\chi_x^2+\chi_y^2}=0.
\end{equation}
Such gap-closing points or critical points, across which the winding number changes, mark the phase boundaries between gapped phases in the phase diagram. Therefore, a transition between the gapped phases occurs with bulk gap closing and opening along with a change in the number of edge modes. 

\begin{figure}[t]
	\includegraphics[width=4.2cm,height=3cm]{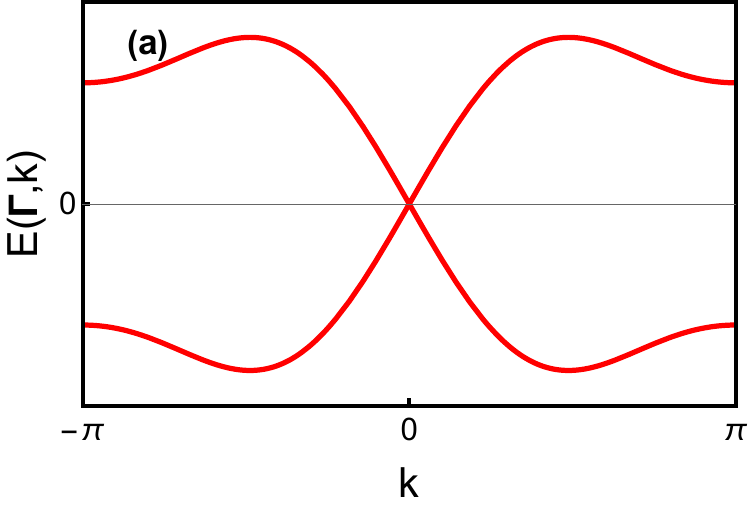}\includegraphics[width=4.2cm,height=3cm]{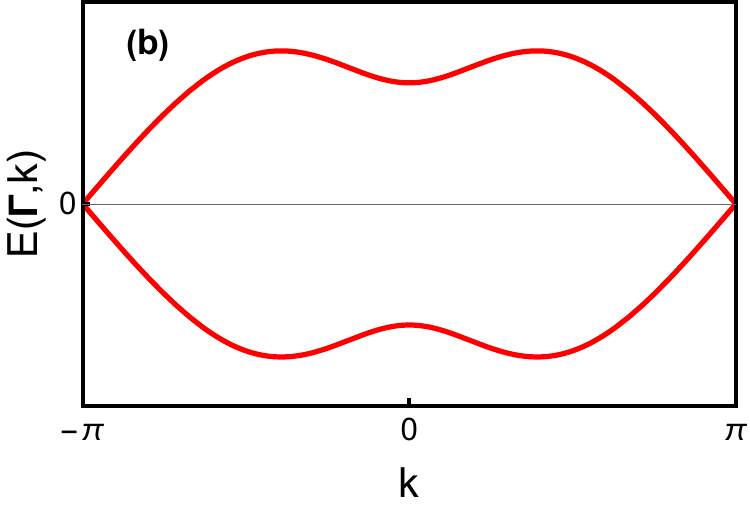}\\
    \includegraphics[width=4.2cm,height=3cm]{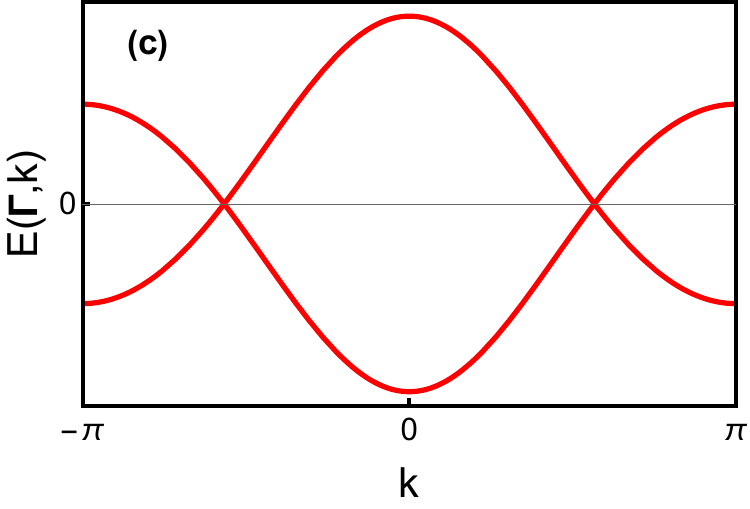}\includegraphics[width=4.2cm,height=3cm]{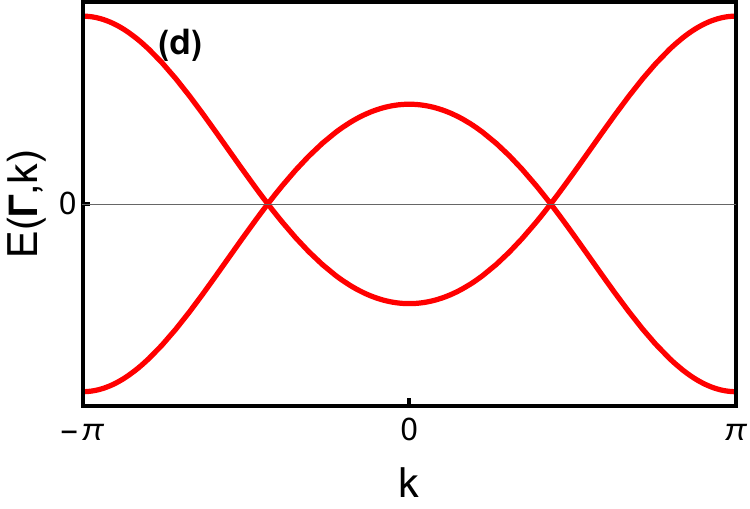}\\
    \includegraphics[width=4.2cm,height=3cm]{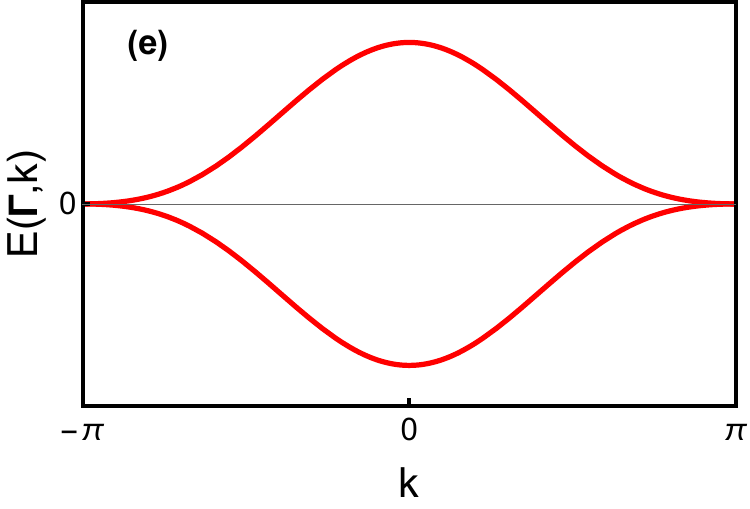}\includegraphics[width=4.2cm,height=3cm]{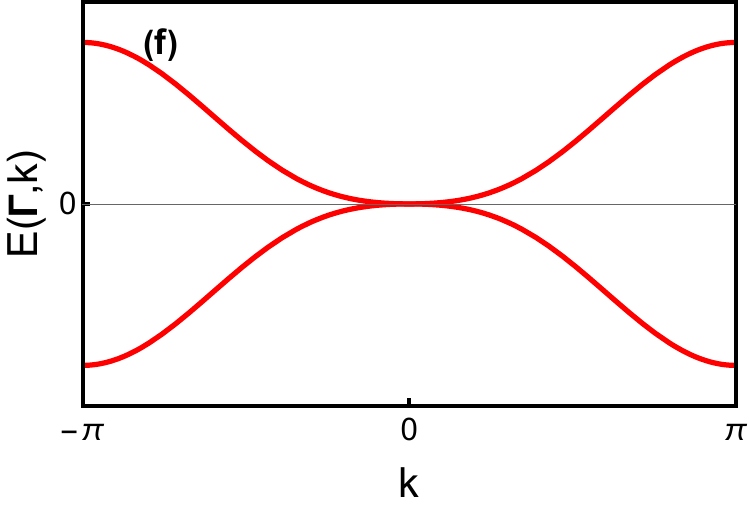}\\
    \includegraphics[width=4.2cm,height=3cm]{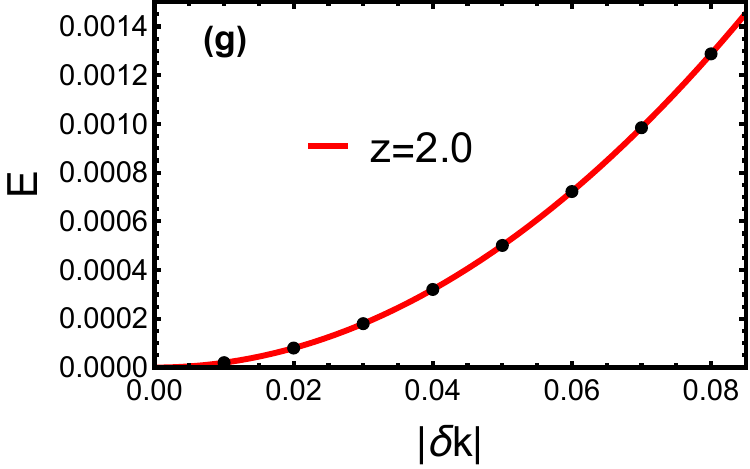}\includegraphics[width=4.2cm,height=3cm]{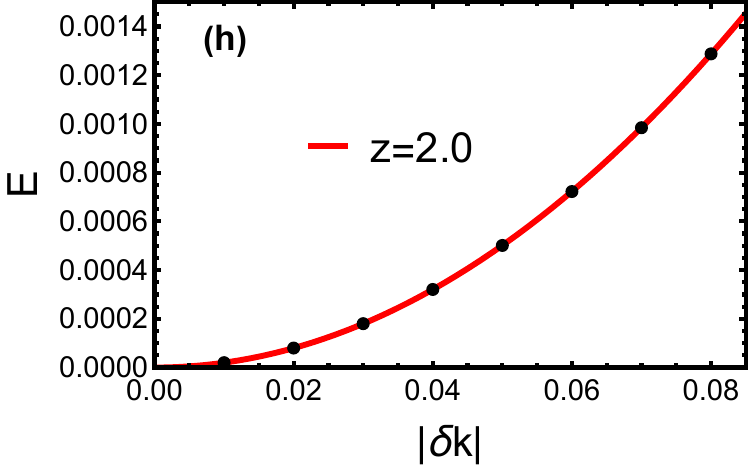}\\
	\caption{Dispersion at critical and multicritical points shown in Fig.\ref{fig1} (for $\Gamma_{0}=0.2$ and $\Gamma_{2}=1$). (a) At $\Gamma_1=1$ and $\Gamma_3=-2.2$ on the red critical line, where the gap closes at $k_0=0$. (b) At $\Gamma_1=2$ and $\Gamma_3=-0.8$ on the blue critical line, where the gap closes at $k_0=\pm\pi$. (c) At $\Gamma_1=2$ and $\Gamma_3=1.92$ on the magenta critical line, where the gap closes at $k_0=\pm 0.56\pi$. (d) At $\Gamma_1=-2$ and $\Gamma_3=-1.92$ on the green critical line, where the gap closes at $k_0=\pm 0.43\pi$. (e) At $\Gamma_1=0.8$ and $\Gamma_3=0.4$, i.e., the multicritical point $M_1$ where the gap closes at $k_0=\pm\pi$. (f) At $\Gamma_1=-0.8$ and $\Gamma_3=-0.4$, i.e., the multicritical point $M_2$, where the gap closes at $k_0=0$. (g,h) The quadratic nature of the dispersion, at the multicritical points in (e) and (f), is captured using the dynamical critical exponent $z=2$.}
	\label{fdisp}
\end{figure}

The model hosts three topologically distinct gapped phases with $w=0,1,2,3$. Under open boundary conditions, $w=1,2,3$ phases correspond to the open chain with one, two, and three edge modes at each end, respectively, while $w=0$ corresponds to the topologically trivial phase with no edge modes. As shown in Fig.~\ref{fig1}, the gapped phases with $w=1,2,3$ are separated by the lines of critical points, throughout which the bulk gap is closed. Thus, the critical lines set the phase boundaries between various gapped phases in the parameter space and are marked in different colors as shown in Fig.~\ref{fig1}. The red and blue lines are the high symmetry critical lines where the gap closing occurs at critical momenta $k_0=0$ and $k_0=\pm \pi$ respectively, as shown in Fig.~\ref{fdisp}(a) and (b). The magenta and green critical lines are non-high symmetry critical lines where gap closing occurs for arbitrary values of $k_0$ which depends on parameter space as $k_0=\pm\arccos((-2\Gamma_{2}+ \sqrt{4\Gamma_{2}^2-16\Gamma_{3}(\Gamma_{1}-\Gamma_{3})})/8\Gamma_{3})$ and $k_0=\pm\arccos((-2\Gamma_{2}- \sqrt{4\Gamma_{2}^2-16\Gamma_{3}(\Gamma_{1}-\Gamma_{3})})/8\Gamma_{3})$ respectively, as shown in Fig.~\ref{fdisp}(c) and (d). These critical momenta yield the critical lines
\begin{align}
C_1:\Gamma_3&=-(\Gamma_0+\Gamma_1+\Gamma_2)\label{r}\\
C_2:\Gamma_3&=(\Gamma_0-\Gamma_1+\Gamma_2)\label{b}\\
C_3:\Gamma_{3}&=\left( \frac{\Gamma_{1}+ \sqrt{\Gamma_{1}^2+4\Gamma_{0}(\Gamma_{0}-\Gamma_{2})}}{2}\right) \label{m}\\
C_4:\Gamma_{3}&=\left( \frac{\Gamma_{1}- \sqrt{\Gamma_{1}^2+4\Gamma_{0}(\Gamma_{0}-\Gamma_{2})}}{2}\right)\label{g}
\end{align}
where $C_1$, $C_2$, $C_3$, and $C_4$ represent the critical lines that are red, blue, magenta, and green lines, respectively, in Fig.~\ref{fig1}. 
In the parameter space, the region defined by 
$\Gamma_3<C_1$, $\Gamma_3>C_2$, $\Gamma_3>C_3$, and $\Gamma_3<C_4$ corresponds to a topological gapped phase with $w=3$. 
The region defined by 
$\Gamma_3<C_1$, $\Gamma_3>C_2$, $\Gamma_3<C_3$, and $\Gamma_3>C_4$ 
corresponds to a topological gapped phase with $w=1$. 
Similarly, the region defined by 
$\Gamma_3>C_1$ and $\Gamma_3<C_2$
corresponds to a topological gapped phase with $w=2$, as shown in Fig.~\ref{fig1}. 

Note that, for fixed $\Gamma_2=1$, the Eq.\ref{r} to Eq.\ref{g} are critical surfaces separating gapped topological phases in three-dimensional parameter space. We also fix $\Gamma_0=0.2$, hence the critical surfaces are called critical lines. The $\Gamma_0$ is fixed such that the parameter space is completely non-trivial (has only gapped phases with non-zero $w$) as shown in Fig.~\ref{fig1}. For a higher value of $\Gamma_0$ we recover a trivial gapped phase with $w=0$, which is discussed in Ref.\cite{kumar2023topological}. Therefore, a condition can be obtained as $\Gamma_0=\Gamma_2/3$, such that for $\Gamma_0>\Gamma_2/3$ the parameter space possesses a trivial $w=0$ phase and for $\Gamma_0\leq\Gamma_2/3$, the $w=0$ phase vanishes, leaving a non-trivial parameter space. The detailed reasoning to obtain this condition is discussed in Section \ref{sec4}.

In general, instances where more than two gapped phases meet at a critical points gives rise to multicritical behavior at that point. Moreover, these multicritical points are also identified at the intersection of two or more critical lines. 
In our model, such multicritical points occur at the intersection of high and non-high symmetry critical lines. As shown in Fig.~\ref{fig1}, blue and magenta lines intersect at a multicritical point $M_1$ (orange dot) obtained at $\Gamma_{1}=(3\Gamma_{0}+\Gamma_{2})/2$. Similarly, red and green lines intersect at $M_2$ (purple dot) obtained at $\Gamma_{1}=-(3\Gamma_{0}+\Gamma_{2})/2$. Both of these multicritical points are characterized by the quadratic dispersions at the gap closing points $k_0=\pm \pi$ for $M_1$ and $k_0=0$ for $M_2$, respectively, as shown in Fig.~\ref{fdisp}(e) and (f). The nature of dispersion near a gap closing momenta $k_0$ can be identified from the dynamical critical exponent $z$ defined as $E(\boldsymbol{\Gamma},k)\propto k^z$. At $M_{1,2}$ we get exponents to be $z=2$, as shown in the Fig.~\ref{fdisp}(g) and (h), ensuring the quadratic nature of dispersion near $k_0$. 

As a direct consequence of higher winding number gapped topological phases, the model exhibits topologically distinct critical phases \cite{verresen2018topology,kumar2021topological,kumar2023topological}. Certain critical points exhibit nontrivial localization of the edge modes despite gapless bulk. However, some critical points remain trivial without edge mode localization. Therefore, the topologically trivial and nontrivial critical points define the critical segments, on a critical line, which are separated by a multicritical point. These points act as a topological phase transition point between distinct gapless phases along a critical line. Unlike the conventional case, this transition does not involve bulk gap opening and closing during phase transition via multicriticalities. Moreover, it has been previously observed that the multicriticalities pin the phase boundary between critical phases at which edge modes completely delocalize into bulk \cite{kumar2021topological,kumar2023topological}.

In our model, we find such nontrivial critical phases on both high-symmetry and non-high-symmetry critical lines. There are two distinct high-symmetry critical phases and one non-high-symmetry critical phase. These phases are separated by the multicritical points $M_{1,2}$. 
The high symmetry critical phases that separate the $w=2$ and $w=3$ gapped phases (i.e., $\Gamma_1>M_1$ and $\Gamma_1<M_2$ on blue and red critical lines in Fig.~\ref{fig1} respectively) are non-trivial and host two edge modes at each end of the open chain. Similarly, the high symmetry critical phases that separate the $w=2$ and $w=1$ gapped phases (i.e., $\Gamma_1<M_1$ and $\Gamma_1>M_2$ on blue and red critical lines respectively) and non-high symmetry critical phases that separate the $w=3$ and $w=1$ gapped phases (i.e., $\Gamma_1>M_1$ and $\Gamma_1<M_2$ on magenta and green critical lines respectively) both host one edge mode at each end of the chain.
We refer to Appendix~\ref{ED-Crit} for open boundary solutions at these critical phases. Therefore, in our model, all the critical phases show topological non-trivial characteristics. 

\begin{figure}[t]
	\includegraphics[width=4.2cm,height=2.8cm]{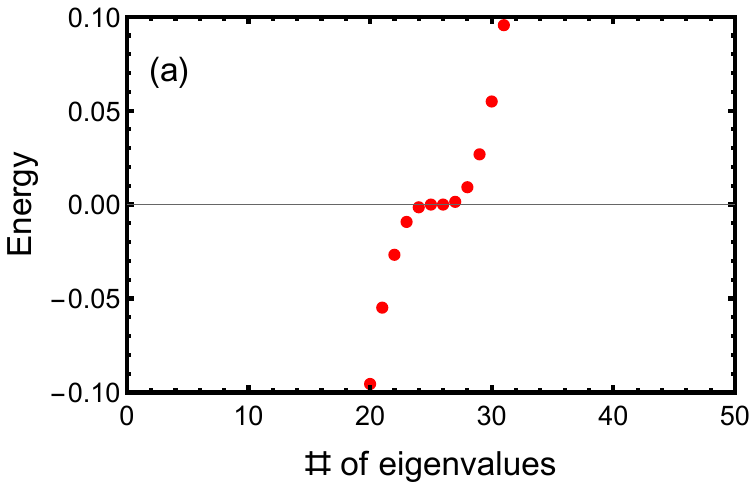}\hspace{0.2cm}\includegraphics[width=4.2cm,height=2.8cm]{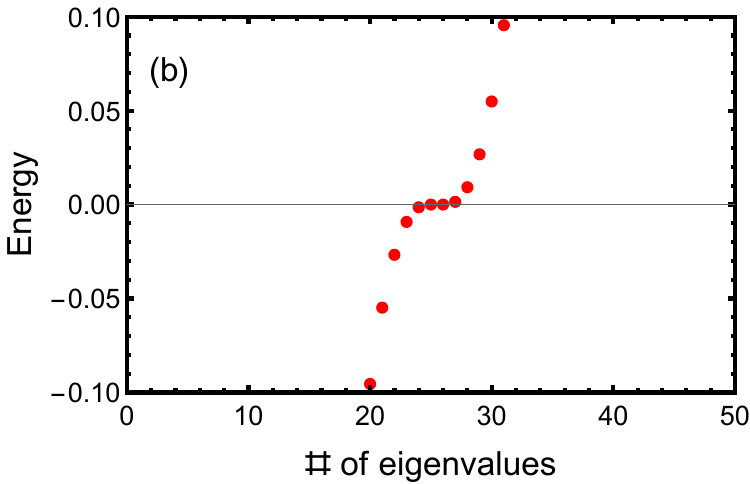}
	\caption{Eigenvalue distribution under open boundary condition at multicritical points. The eigenvalues near the zero energy are shown for $\Gamma_{0}=0.2$ and $\Gamma_{2}=1$. (a) For $M_1$ at $(\Gamma_{1},\Gamma_{3})=(0.8,0.4)$. (b) For $M_2$ at $(\Gamma_{1},\Gamma_{3})=(-0.8,-0.4)$. Both multicriticalities show one pair of eigenvalues that are localized at zero energy.}
	\label{EDM}
\end{figure}

\section{Topological invariant}\label{sec3}
The characterization of the topological properties of the critical phases is tricky since the bulk has no gapped degrees of freedom. Therefore, the conventional bulk-boundary correspondence fails \cite{verresen2018topology} and the topological invariant number in Eq.\ref{windnum} is ill-defined at a critical point. This is because the curvature function diverges (i.e. $\chi^2_x(\boldsymbol{\Gamma},k)+\chi^2_y(\boldsymbol{\Gamma},k)=0$) at a critical point. Nonetheless, an approach to define the topological invariant, by avoiding an exact critical point and recovering the number of edge modes from it, has been proposed \cite{kumar2021topological,verresen2020topology}. However, such a topological invariant contains the information of both critical and topological properties encapsulated within. Therefore, an approximation scheme has to be implemented to extract the information of only the number of edge modes that can appear. 

Alternatively, a simple way of obtaining the topological invariant to count the number of edge modes and the localization properties is shown in Ref.~\cite{verresen2018topology}. This method is based on the zeros and poles of a complex function associated with the Hamiltonian. The topological invariant obtained from this method can effectively characterize both gapped and critical phases of the model. At first, by writing fermionic operators into Majorana operators followed by a Fourier transformation, a complex function $f(\zeta)$ where $\zeta=e^{ik}$ is a complex number, can be obtained. This dictates the $f(\zeta)$ on the unit circle as $k$ varies
over the Brillouin zone.  According to Cauchy’s argument principle, the topological invariant can be obtained as $w_c=N_z-N_p$ where $N_z$ and $N_p$ are zeros and poles of the complex function within the unit circle. The topological invariant $w_c$ can efficiently count the localized edge modes at a critical phase. Moreover, the localization lengths of the edge modes can be obtained using the zeros as $\xi=-1/\ln(|\zeta|)$.

Tuning the parameters into different gapped phases drives the zeros in and out of the unit circle. Therefore, at a critical point, one or more zeros lie on the unit circle. If the zeros on the unit circle are non-degenerate, the corresponding conformal field theory (CFT) of the critical phase can be measured in terms of the central charge $c$. Therefore, the non-degenerate zeros on the unit circle determine the central charge and are given by $c=(1/2)N^{\prime}_z$, where $N^{\prime}_z$ is the number of zeros on the unit circle in the complex plane. The high symmetry and non-high symmetry critical phases of our model are characterized by the value of $c$, which enables the distinction between phases based on their critical properties apart from their topological features. 

\begin{figure}[t]
	\includegraphics[width=4.2cm,height=3.0cm]{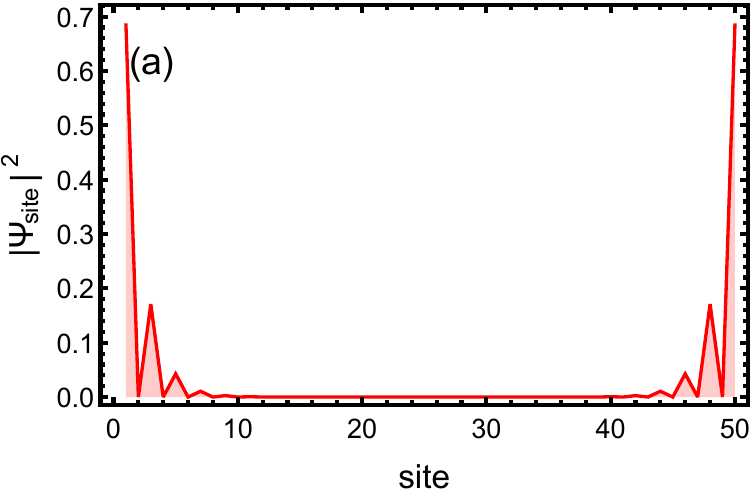}\includegraphics[width=4.2cm,height=3.0cm]{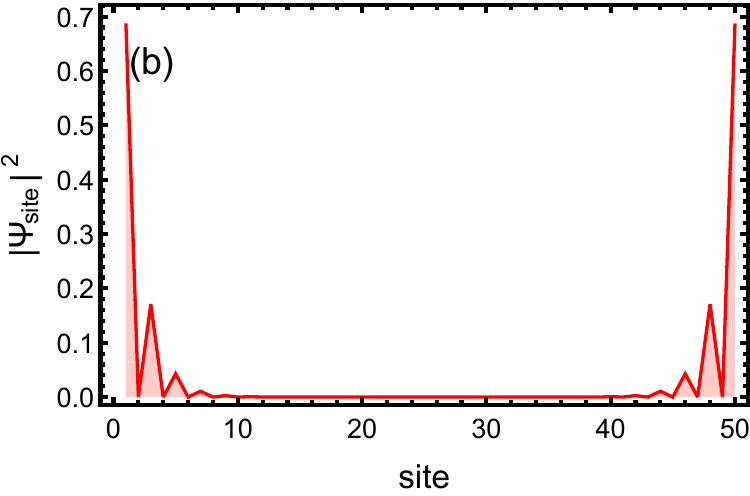}
	\caption{Probability distribution at topological multicritical points for $\Gamma_{0}=0.2$ and $\Gamma_{2}=1$. (a) For $M_1$ at $(\Gamma_{1},\Gamma_{3})=(0.8,0.4)$. (b) For $M_2$ at $(\Gamma_{1},\Gamma_{3})=(-0.8,-0.4)$. The distribution at the edge indicates the topological nontriviality of the multicritical points.}
	\label{PDM}
\end{figure}

The complex function associated with our model Hamiltonian can be obtained as 
\begin{equation}
f(\zeta)=-\frac{\Gamma_{0}}{2}-\frac{\Gamma_{1}}{2}\zeta -\frac{\Gamma_{2}}{2} \zeta^2-\frac{\Gamma_{3}}{2} \zeta^3. \label{complexx}
\end{equation}
with the zeros
\begin{align}
\zeta_1&=-\frac{\sqrt[3]{\kappa}}{3 \sqrt[3]{2} \text{$\Gamma_3 $}}
+\frac{\sqrt[3]{2} \left(3 \text{$\Gamma_1 $} \text{$\Gamma_3 $}-\text{$\Gamma_2 $}^2\right)}{3 \text{$\Gamma_3 $} \sqrt[3]{\kappa}}
-\frac{\text{$\Gamma_2 $}}{3 \text{$\Gamma_3 $}} \label{zero1}\\
\zeta_2&=-\frac{\left(1+i \sqrt{3}\right) \left(3 \text{$\Gamma_1 $} \text{$\Gamma_3 $}-\text{$\Gamma_2 $}^2\right)}{3\ 2^{\frac{1}{3}} \text{$\Gamma_3 $} \sqrt[3]{\kappa}}
+\frac{\left(1-i \sqrt{3}\right) \sqrt[3]{\kappa}}{6 \sqrt[3]{2} \text{$\Gamma_3 $}}
-\frac{\text{$\Gamma_2 $}}{3 \text{$\Gamma_3 $}} \label{zero2}\\
\zeta_3&=\frac{\left(1+i \sqrt{3}\right) \sqrt[3]{\kappa}}{6 \sqrt[3]{2} \text{$\Gamma_3 $}}
-\frac{\left(1-i \sqrt{3}\right) \left(3 \text{$\Gamma_1 $} \text{$\Gamma_3 $}-\text{$\Gamma_2 $}^2\right)}{3\ 2^{2/3} \text{$\Gamma_3 $} \sqrt[3]{\kappa}}
-\frac{\text{$\Gamma_2 $}}{3 \text{$\Gamma_3 $}}\label{zero3}
\end{align}
where 
\begin{align*}
\kappa&=\sqrt{\left(27 \Gamma_0 \Gamma_3^2-9 \Gamma_1 \Gamma_2 \Gamma_3+2 \Gamma_2^3\right)^2+4 \left(3 \Gamma_1 \Gamma_3-\Gamma_2^2\right)^3} \\
&+27 \Gamma_0 \Gamma_3^2-9 \Gamma_1 \Gamma_2 \Gamma_3+2 \Gamma_2^3
\end{align*}
The complex function $f(\zeta)$ does not have poles ($N_p=0$) and, therefore, the number of zeros inside the unit circle gives the winding number of the corresponding critical phases. We find that the distinct phases on the high symmetry critical lines are identified with $w_c=2$ and $w_c=1$, along with the central charge $c=1/2$. The critical phases on the non-high symmetry lines are $w_c=1$ phases with $c=1$. We refer to Appendix.\ref{W-crit} for the details of zeros at the various critical phases. The value of $c$ identifies the high symmetry and non-high symmetry criticalities. The value of $w_c$ is consistent with the edge mode solutions obtained for corresponding critical phases and indicates that the model possesses only topologically nontrivial critical phases, as shown in Fig.\ref{fig1}.

\begin{figure}[t]
	\includegraphics[width=3.3cm,height=3.0cm]{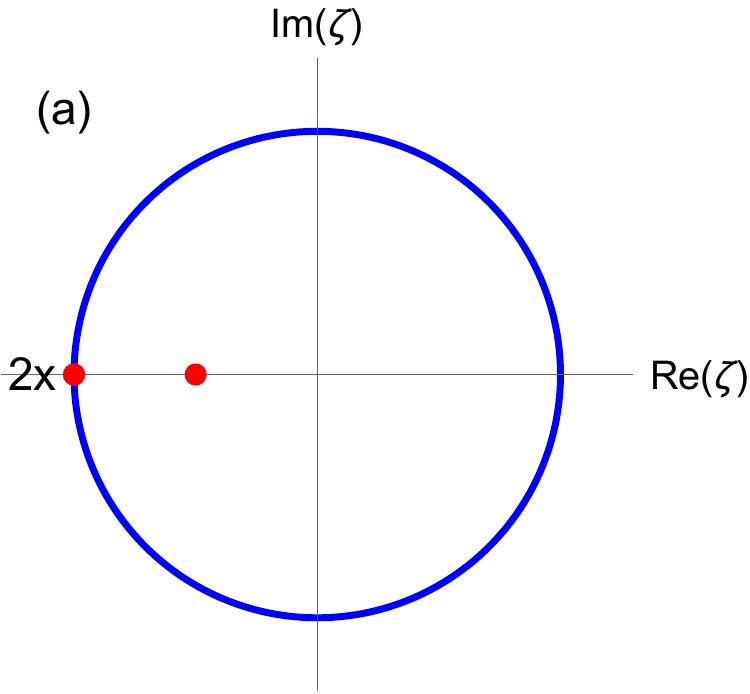}\hspace{0.2cm}\includegraphics[width=3.3cm,height=3.0cm]{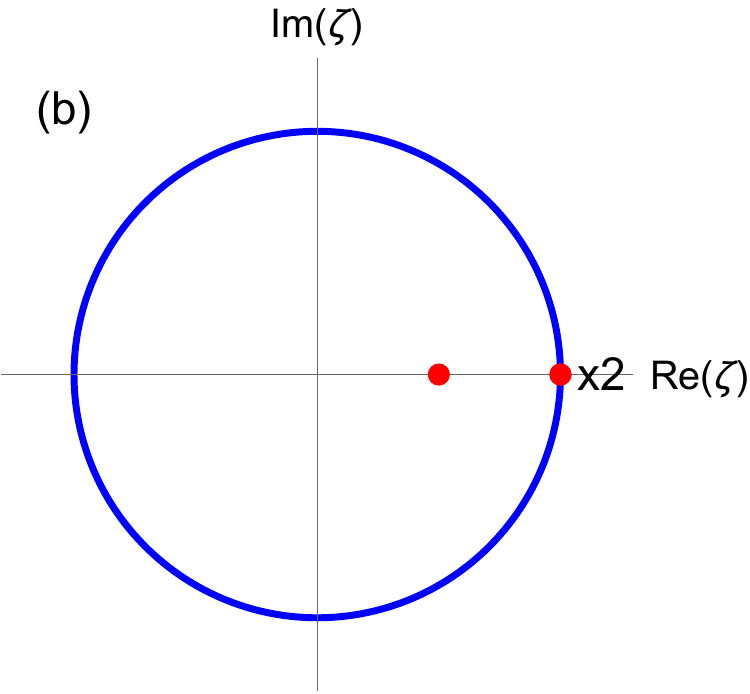}
	\caption{Zeros of the complex function on a complex plane at multicritical points for $\Gamma_{0}=0.2$ and $\Gamma_{2}=1$. (a) For $M_1$ at $(\Gamma_{1},\Gamma_{3})=(0.8,0.4)$. (b) For $M_2$ at $(\Gamma_{1},\Gamma_{3})=(-0.8,-0.4)$. Both multicriticalities show one zero inside and two degenerate zeros on the unit circle, leading to $w_c=1$.}
	\label{ZCM}
\end{figure}

Tuning the parameters on the critical lines, the transition between topological critical phases can be observed. This topological transition occurs at the multicritical points. The critical phases $w_c=2$ and $w_c=1$ undergo a topological phase transition between them at the points $M_{1,2}$. Note that $w_c=1$ is obtained for both high and non-high symmetry critical phases. 
Therefore, since these two phases are topologically equivalent (having same topological invariant), there is no topological transitions between them at the multicritical points, i.e., the number of edge modes remains the same.
However, the transition between these two phases involves a change in $c$ value, which represents a phase transition between distinct CFTs at the multicritical points. Moreover, the nature of the dispersion near the gap closing momenta $k_0$ changes at the transition point and remains the same at the critical phases. As a consequence, we have dynamical exponent $z=1$ at the critical phases and $z=2$ at $M_{1,2}$. 

In previous studies \cite{kumar2021topological,kumar2023topological}, it has been observed that the localization length of the edge modes at the critical phases increases as the parameters are tuned towards the multicritical points. Therefore, in such cases, the edge modes completely delocalize into the bulk at the exact multicritical points, making them topologically trivial. However, in the following Section, we show that in our model the edge mode localization can also occur at the multicritical points, leading to topologically nontrivial multicriticalities.

\begin{figure}[t]
	\includegraphics[width=4.3cm,height=3.1cm]{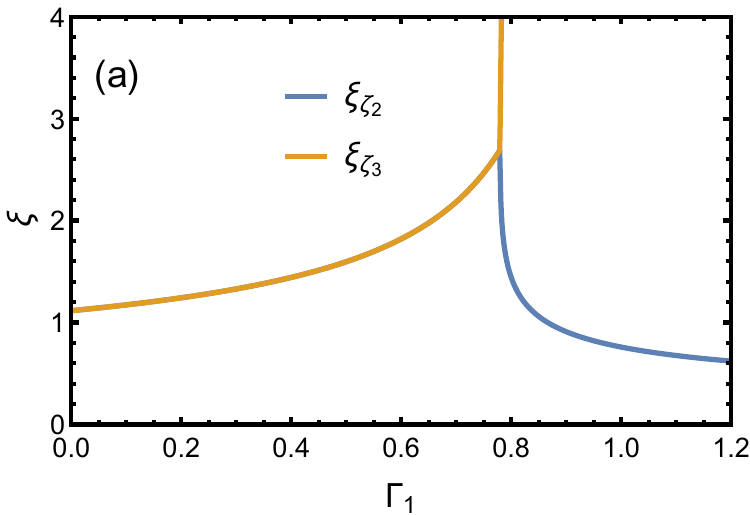}\includegraphics[width=4.3cm,height=3.1cm]{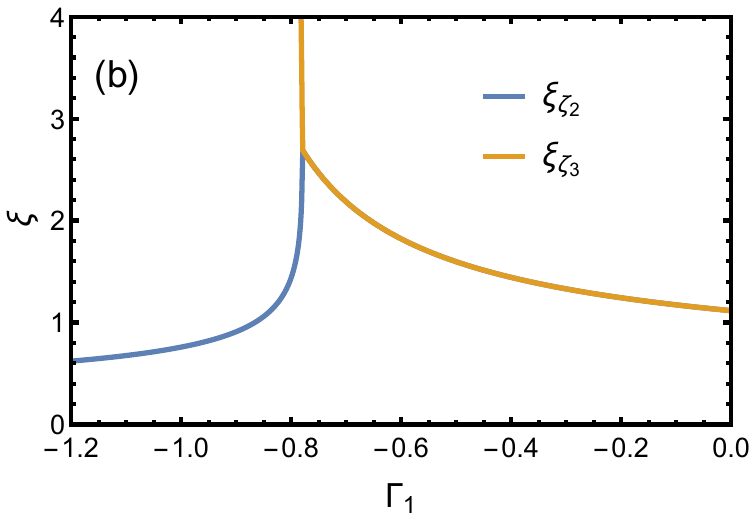}
	\caption{Localization lengths of edge modes, along the high symmetry critical lines, calculated from the zeros with the parameters $\Gamma_{0}=0.2$ and $\Gamma_{2}=1$. (a) For $M_1$ at $(\Gamma_{1},\Gamma_{3})=(0.8,0.4)$. (b) For $M_2$ at $(\Gamma_{1},\Gamma_{3})=(-0.8,-0.4)$. Edge modes with $\xi_{\zeta_2}$ remain localized at the multicritical points while edge modes with $\xi_{\zeta_3}$ delocalize into the bulk.}
	\label{LL}
\end{figure}

\section{Topological Multicritical Points}\label{sec4}
The topological characters of the multicritical points can be identified using eigenvalue distribution under open boundary conditions. The multicritical points $M_1$ and $M_2$ in Fig.~\ref{fig1} are at $(\Gamma_{1},\Gamma_{3})=(0.8,0.4)$ and $(\Gamma_{1},\Gamma_{3})=(-0.8,-0.4)$ respectively. Eigenvalue distribution at both $M_{1,2}$, as shown in Fig.~\ref{EDM}, indicates the presence of one pair of eigenvalues at zero energy that corresponds to one edge mode localized at each end of the open chain of size $N=25$. 
Moreover, the nontriviality of the multicritical points can be identified using a corresponding probability distribution. Fig.~\ref{PDM} shows the distribution of probability density in an open chain. At both $M_{1,2}$, we observe the distribution is dominant at the edge of the chain, representing localized edge modes at the multicritical points. 

These edge modes at multicriticalities can also be characterized by analyzing the distribution of zeros in the complex plane of the complex function associated with the Hamiltonian. At $M_1$, we observe among the three roots $\zeta_1=\zeta_3=-1$ and $\zeta_2<1$, as shown in Fig.~\ref{ZCM}(a). Therefore, at this point, we get the topological invariant at multicritical point $w_{mc}=1$, as there is only one zero lying inside the unit circle, which indicates one localized edge mode. Similarly, at $M_2$ we observe $\zeta_1=\zeta_3=1$ and $\zeta_2<1$ which yields $w_{mc}=1$ indicating one localized edge mode, as shown in Fig.~\ref{ZCM}(b). These results are consistent with the eigenvalue and probability distributions under open boundary conditions. Moreover, both $M_{1,2}$ hosts degenerate zeros $\zeta_{1,3}$ 
and therefore the bulk is not described by a CFT \cite{verresen2018topology}. 

\begin{figure}[t]
	\includegraphics[width=7cm,height=10.3cm]{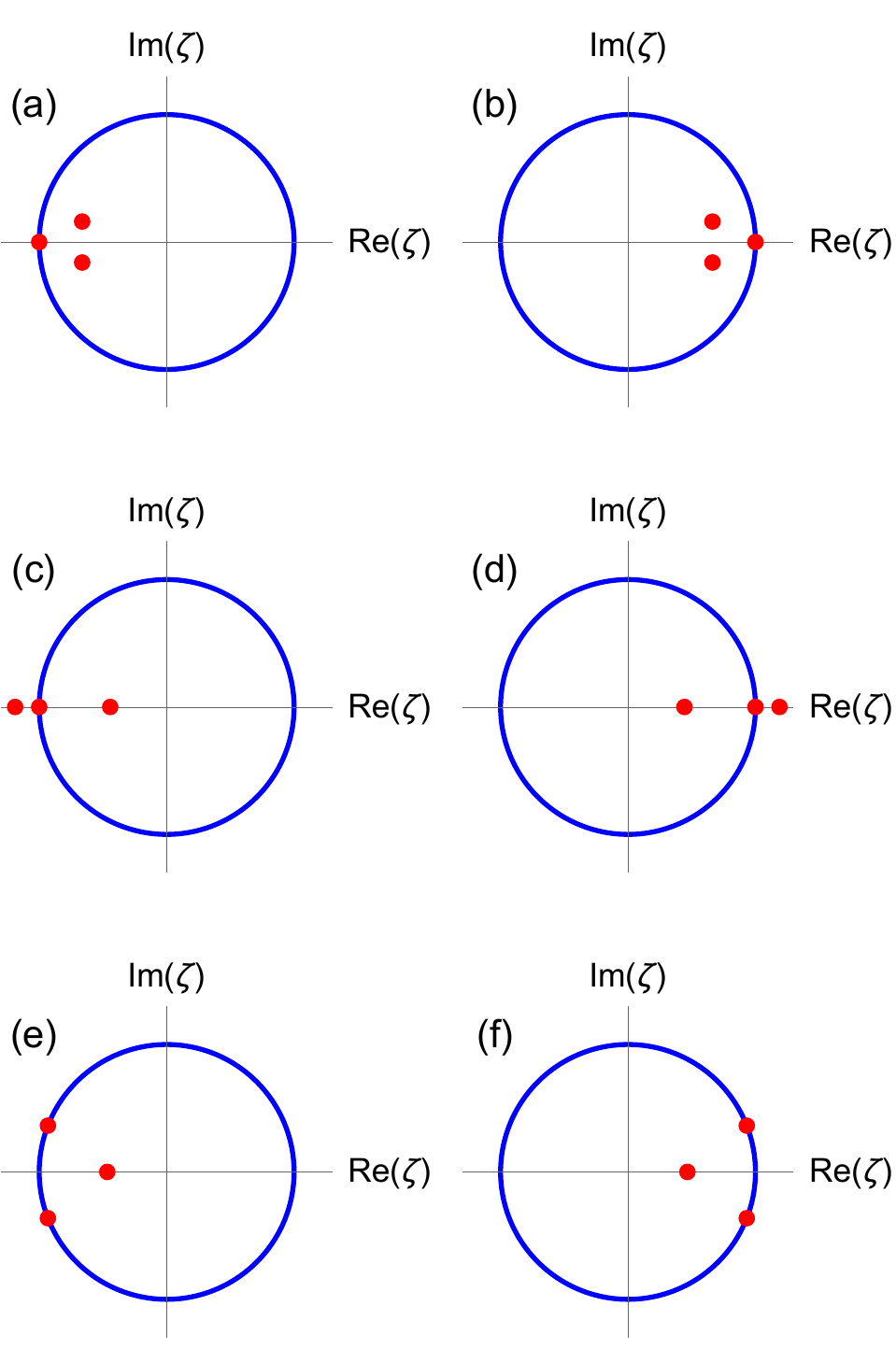}
	\caption{Zeros at the critical points close to the multicritical points $M_1$ [(a),(c),(e)] and $M_2$ [(b),(d),(f)] for parameters $\Gamma_{0}=0.2$ and $\Gamma_{2}=1$. At $w_c=2$, on (a) blue [at $(\Gamma_{1},\Gamma_{3})=(0.77,0.43)$] and (b) red [at $(\Gamma_{1},\Gamma_{3})=(-0.77,-0.43)$] critical lines, we get $\zeta_1=\pm 1$ and $\zeta_{2,3}<\pm 1$. At $w_c=1$, on (c) blue [at $(\Gamma_{1},\Gamma_{3})=(0.82,0.38)$] and (d) red [at $(\Gamma_{1},\Gamma_{3})=(-0.82,-0.38)$] critical lines, we get $\zeta_1>\pm 1$, $\zeta_2<\pm 1$ and $\zeta_3=\pm 1$. At $w_c=1$, on (e) magenta [at $(\Gamma_{1},\Gamma_{3})=(0.802,0.43)$] and (f) green [at $(\Gamma_{1},\Gamma_{3})=(-0.802,-0.43)$] critical lines, we get $\zeta_1<\pm1$ and $\zeta_{2,3}$ on the unit circle.}
	\label{ZNM}
\end{figure}

\begin{figure}[t]
	\includegraphics[width=4.3cm,height=3.1cm]{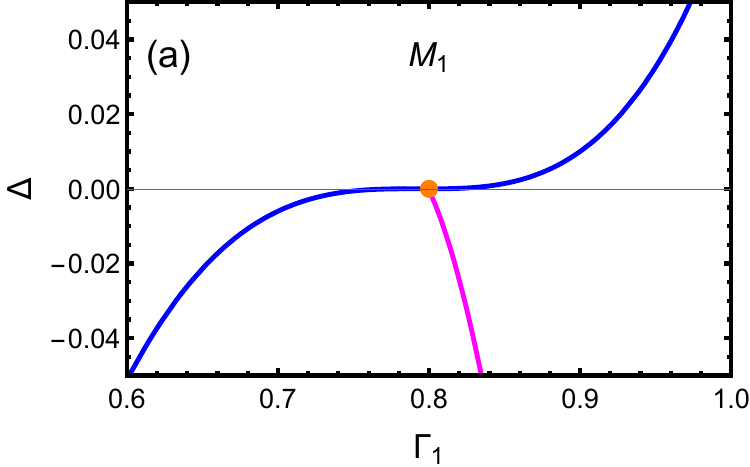}\includegraphics[width=4.3cm,height=3.1cm]{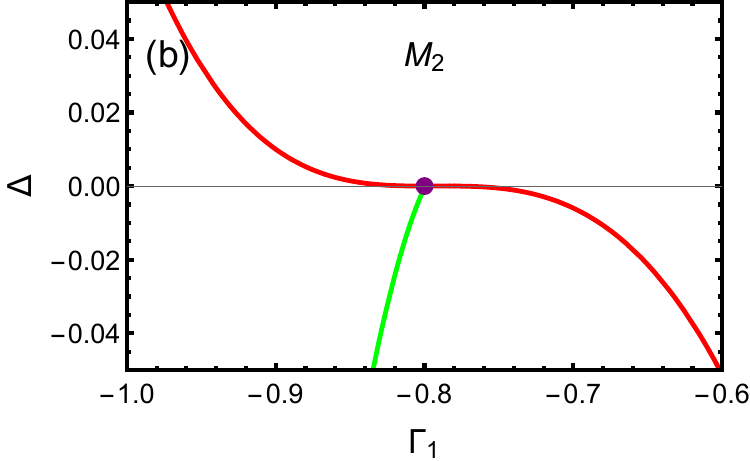}
	\caption{Discriminant in the vicinity of topological multicritical points. The plots are obtained for $\Gamma_{0}=0.2$ and $\Gamma_{2}=1$ along the corresponding critical lines. (a) For $M_1$ at $\Gamma_{1}=0.8$. Discriminant along the blue critical line is negative at $w_c=2$ ($\Gamma_1<0.8$) and positive at $w_c=1$ ($\Gamma_1>0.8$) critical phases. Along the magenta critical line, i.e., at $w_c=1$ phase, we get $\Delta<0$. (b) For $M_2$ at $\Gamma_{1}=-0.8$. Discriminant along the red critical line is negative for at $w_c=2$ ($\Gamma_1>-0.8$) and positive at $w_c=1$ ($\Gamma_1<-0.8$) critical phases. Along the green critical line, i.e., at $w_c=1$ phase, we get $\Delta<0$. The discriminant $\Delta=0$ at both the multicritical points and flips sign on either side.}
	\label{Dis}
\end{figure}

We also calculate the localization length of the edge modes using zeros of the complex function along the high symmetry critical lines (blue and red lines). 
The localization length can be obtained as $\xi_{\zeta_i}=-1/\ln(|\zeta_i|)$, where the zeros that fall inside the unit circle give finite positive localization lengths of the stable edge modes. As shown in Fig.~\ref{LL}(a), along the blue critical line for $w_c=2$ phase we observe two edge modes with same localization lengths $\xi_{\zeta_2}=\xi_{\zeta_3}$ due to complex conjugate pairs ($|\zeta_2|=|\zeta_3|$) inside the unite circle. One of them ($\xi_{\zeta_3}$) diverges at the multicritical point $M_1$ ($\Gamma_{1}=0.8$), indicating delocalization of the corresponding edge mode into the bulk. However, other edge modes remain stable with localization length $\xi_{\zeta_2}$ at multicritical point, leading to $w_{mc}=1$, and even further at $w_c=1$ critical phase. Similar results can be observed for the multicritical point $M_2$ ($\Gamma_{1}=-0.8$), as shown in Fig.~\ref{LL}(b). Therefore, both $M_{1,2}$ are identified as topological multicritical points with one stable localized edge mode at each end of the open chain and are characterized by $w_{mc}=1$. 

Further, we analyze the nature of zeros in terms of the associated discriminant, in the vicinity of multicritical points. The complex function $f(\zeta)$ is a cubic polynomial with roots $\zeta_{1,2,3}$, which are real or complex conjugate pairs.
The discriminant of the polynomial can be obtained using the roots as
\begin{equation}
\Delta=\Gamma_{3}^4 (\zeta_1-\zeta_2)^2 (\zeta_1-\zeta_3)^2 (\zeta_2-\zeta_3)^2
\end{equation}
The sign of the discriminant dictates the nature of the roots on the complex plane. If $\Delta<0$ the polynomial has one real and two complex conjugate roots. If $\Delta>0$, all the roots are real. If $\Delta=0$, the polynomial has degenerate zeros. In the vicinity of $M_1$ ($M_2$), we have $w_c=1,2$ critical phases on the blue (red) line and $w_c=1$ critical phase on the magenta (green) line. Close to the multicriticalities, at $w_c=2$ phase we observe complex conjugate pairs $\zeta_{2,3}$ that lie inside the unit circle while a real root $\zeta_1$ pinned at $\pm1$, as shown in Fig.~\ref{ZNM}(a) and (b). As we tune the parameter space towards the multicriticalities, $\zeta_{2,3}$ slides towards the real axis in the complex plane and finally causes degenerate zeros $\zeta_1=\zeta_3$ at $\pm1$ and $\zeta_2<\pm1$ that lie on the real axis, as shown in Fig.~\ref{ZCM}. Similarly, at $w_c=1$ phase on blue or red critical lines, we observe that all the roots are real. Near the multicriticalities, we get $\zeta_1>\pm1$, $\zeta_2<\pm1$ and $\zeta_3=\pm1$ as shown in Fig.~\ref{ZNM}(c) and (d). As the parameters are tuned, $\zeta_1$ approaches $\pm1$, finally leading to degenerate zeros at 
the multicriticalities with $\zeta_1=\zeta_3$ while keeping $\zeta_2<\pm1$. On the other hand, at the critical phases $w_c=1$ on the magenta and green lines close to the multicriticalities, we observe complex conjugate pairs $\zeta_{2,3}$ on the unit circle while a real root $\zeta_1$ lie inside, as shown in the Fig.~\ref{ZNM}(e) and (f). Tuning the parameters leads to the confluence of complex conjugate pairs towards the real axis with $\zeta_1=\zeta_3$ and $\zeta_2<\pm1$.   

The discriminant for the cases discussed above shows interesting features. Based on the nature of roots at various critical phases, the discriminant is either $\Delta<0$ or $\Delta>0$. At $w_c=2$ (on blue or red lines) and $w_c=1$ (on magenta and green lines) we get $\Delta<0$, as shown in Fig.~\ref{Dis}(a) and (b), since there are a complex conjugate pair and a real root. However, for $w_c=2$ the complex conjugate pair lies inside, while for $w_c=1$ they lie on the unit circle. This distinction marks the different central charges, i.e., $c=1/2$ and $c=1$ respectively, obtained for these phases independently of the topological invariant. At $w_c=1$ (on blue or red lines), we get $\Delta>0$ as all the roots are real. Approaching multicritical points along all these critical phases leads to $\Delta\rightarrow 0$. Moreover, along the high symmetry critical lines, the discriminant flips the sign across the transition between $w_c=2\rightleftarrows w_c=1$ at the multicritical points. Therefore, the topological multicritical points are identified with $\Delta=0$, as shown in Fig.~\ref{Dis}, and they feature positive and negative discriminants on either side. We argue that such behavior is unique to topological multicritical points and is, in general, absent for trivial multicritical points. 

\begin{figure}[t]
	\includegraphics[width=6.2cm,height=4cm]{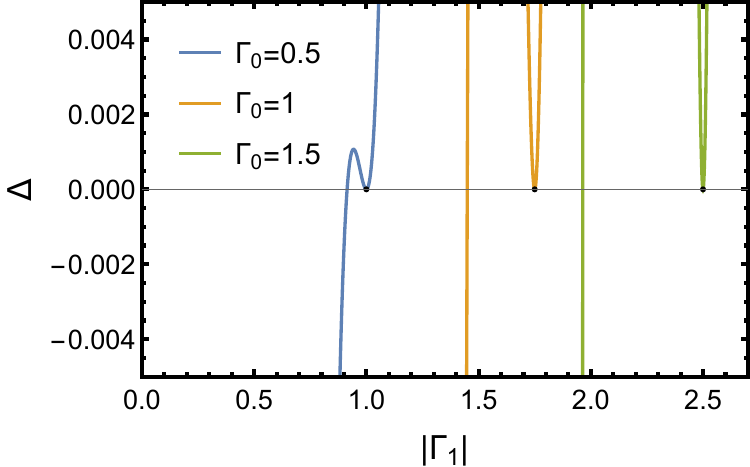}
	\caption{Discriminant in the vicinity of topologically trivial multicritical points. The plots are obtained for fixed $\Gamma_{2}=0.5$ and for $\Gamma_{0}=0.5,1,1.5$ for which the quadratic multicritical points (black dots) are obtained at $\Gamma_{1}=\pm1,\pm1.75,\pm2.5$ respectively, along the corresponding high symmetry critical lines. In all the cases, the discriminant $\Delta=0$ at the multicritical points and remains positive on either side.}
	\label{Dis_tri}
\end{figure}

To show this explicitly, we consider the cases with $\Gamma_{0}>0.2$ such that there can be a trivial phase with $w=0$. In such cases, we look at the behavior of the discriminant at the multicritical points. We consider three cases with $\Gamma_{0}=0.5,1,1.5$ and keep $\Gamma_{2}=0.5$ such that the corresponding topological phase diagram shows the missing (compared to Fig.~\ref{fig1}) $w=0$ gapped phase. One of the above cases is studied in Ref.\cite{kumar2023topological}. In these cases, there are four multicritical points: two with linear and two with quadratic dispersions. Therefore, along a high symmetry critical line, we can witness the transition between $w_c=2\rightleftarrows w_c=0$ at a linear multicritical point and $w_c=0\rightleftarrows w_c=1$ at a quadratic multicritical point.  The edge modes localized at the critical phases delocalize completely into the bulk on approaching these multicriticalities. Therefore, these points are topologically trivial with $w_{mc}=0$. Along the high symmetry critical line for three cases $\Gamma_{0}=0.5,1,1.5$, we obtain the linear multicritical points at $\Gamma_{1}=\pm0.5$ and quadratic points at $\Gamma_{1}=\pm1,\pm1.75,\pm2.5$ respectively. The discriminant at the linear point remains either $\Delta>0$ or $\Delta<0$ with no interesting features associated. 

However, the quadratic points, despite being topologically trivial, exhibit features similar to those in the previous case.
The discriminants for $\Gamma_{0}=0.5,1,1.5$ cases are shown in Fig.~\ref{Dis_tri} where the parameter $|\Gamma_{1}|$ is varied along the high symmetry critical lines in the vicinity of quadratic multicritical points. Clearly, $\Delta=0$ at the quadratic multicritical points obtained at $\Gamma_{1}=\pm1,\pm1.75,\pm2.5$. However, for all these cases $\Delta>0$ at critical phases on either side of the multicritical points and therefore does not show a sign flip as seen in the case of nontrivial multicritical points. This behavior around quadratic multicritical points is due to the presence of a trivial critical phase $w_c=0$. Since this phase does not host any localized edge modes, it leaves the corresponding multicritical points trivial with $w_{mc}=0$. Therefore, a transition from $w_c=0$ to $w_c=1$ brings one of the two roots, which lies outside the unit circle at $w_c=0$ phase, into the unit circle at $w_c=1$ phase along the real axis with degenerate roots ($\pm 1$) at multicritical points. The discriminant thus remains positive on both sides and remains zero at the trivial multicritical points. Note that the discriminant goes to zero also at a critical point away from multicriticality, as seen in Fig.~\ref{Dis_tri}. At these points, the roots are degenerate and they lie outside the unit circle, in contrast to the multicriticalities. Interestingly, these degenerate points come closer to the multicritical points and finally merge at a certain point in the parameter space. This phenomenon eliminates the positive part of the discriminant on one side of the multicritical point, leading to the topological multicritical points with a sign flip of the corresponding discriminants.

The above argument indicates that changing the parameter space changes the number of multicritical points along with their topological properties. Therefore, one can ask what is the criterion for the parameters to obtain topological multicritical points? The presence of trivial gapped and critical phases results in four multicritical points as discussed previously. To obtain a topological phase diagram as in Fig.~\ref{fig1}, where all the trivial gapped and critical phases are removed, can result from the confluence of two kinds (linear and quadratic) of multicriticalities. In the resulting phase diagram, only quadratic multicriticality and nontrivial phases will survive. The multicritical points are obtained at $\Gamma_{1}=\pm(3\Gamma_{0}+\Gamma_{2})/2$ (quadratic) and $\Gamma_{1}=\pm\Gamma_{2}$ (linear). Therefore, a condition for parameters can be obtained as $\Gamma_{0}=\Gamma_{2}/3$ at which the two kinds of multicritical points merge. For $\Gamma_{0}>\Gamma_{2}/3$, trivial phases will survive with trivial multicritical points with $w_{mc}=0$. In contrast, for $\Gamma_{0}< \Gamma_{2}/3$, the parameter space is completely nontrivial with nontrivial multicritical points at which $w_{mc}=1$.

A physically intuitive picture for the emergence of topological character at the quadratic multicritical points can be obtained from the analysis of the asssociated Dirac equation. The kinetic inversion mechanism, discussed previously~\cite{verresen2020topology,kumar2021topological}, manifests through higher-order momentum terms and governs the topological nature of these multicritical points.
	The effective Dirac equation at multicriticality can be obtained by expanding $\chi_x(\boldsymbol{\Gamma}_{mc},k)$ and $\chi_y(\boldsymbol{\Gamma}_{mc},k)$, where $\boldsymbol{\Gamma}_{mc}$ are the parameters at multicritical points, around $k_0$ %(either $0$ or $\pm\pi$) 
	up to fourth order
	\begin{equation}
	H(\boldsymbol{\Gamma}_{mc},k)\approx (\alpha_1k^2+\alpha_2k^4)\sigma_x+\alpha_3k^3\sigma_y
	\end{equation}
	where $\alpha_1=-3\Gamma_0+\Gamma_2$,
	 $\alpha_2=39\Gamma_0-25\Gamma_2$
	 and  $\alpha_3=-12\Gamma_0+6\Gamma_2$.
	 In real space, the zero energy solution, $H\psi(x)=0$, reveals that the spinor $\psi(x)=\rho_{s}\phi(x)$ (where $s=-\text{sign}(\alpha_3/\alpha_2)$) is an eigenstate of $\sigma_z$. Using the trial wavefunction $\phi(x)\propto e^{-x/\xi}$, the decay length of the edge mode can be obtained as $\xi_+\approx |\alpha_3|/\alpha_1$. This highlights that the quadratic kinetic coefficient $\alpha_1$ effectively plays the role of a mass term, 
	 remaining positive (negative) at topological (trivial) phase. We observe that (i) $\alpha_1<0$ for $\Gamma_0>\Gamma_2/3$ (trivial phase with $\xi_+<0$), (ii) $\alpha_1>0$ for $\Gamma_0<\Gamma_2/3$ (non-trivial phase with $\xi_+>0$) and (iii) $\alpha_1=0$ for $\Gamma_0=\Gamma_2/3$. Therefore, multicritical points that are trivial for $\Gamma_0>\Gamma_2/3$, undergo a transition into topological multicritical points at $\Gamma_0=\Gamma_2/3$ and remain non-trivial for $\Gamma_0<\Gamma_2/3$.
	This agrees with the analytical and numerical results discussed before. Therefore, due to the quadratic nature of the multicritical points, higher order kinetic inversion facilitates the emergence of topologically protected edge modes at the multicritical points and dictates the condition for the parameters to obtain topological multicritical points.

\begin{figure*}[t]
	\includegraphics[height=3.57cm]{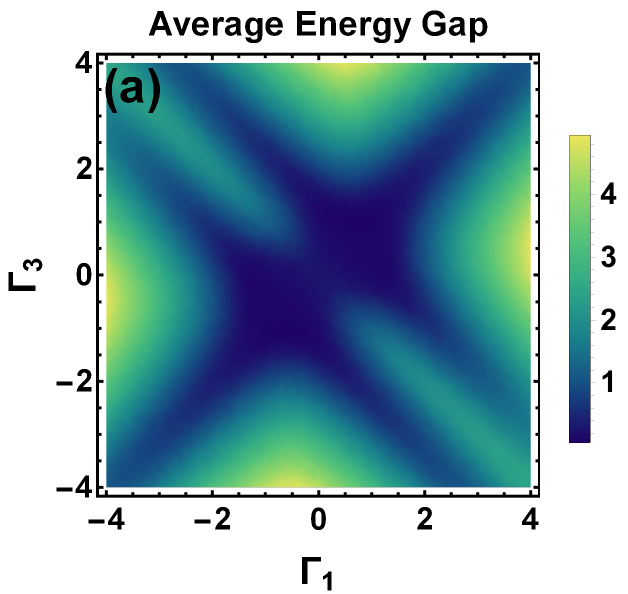}\hspace{0.5cm}
	\includegraphics[height=3.57cm]{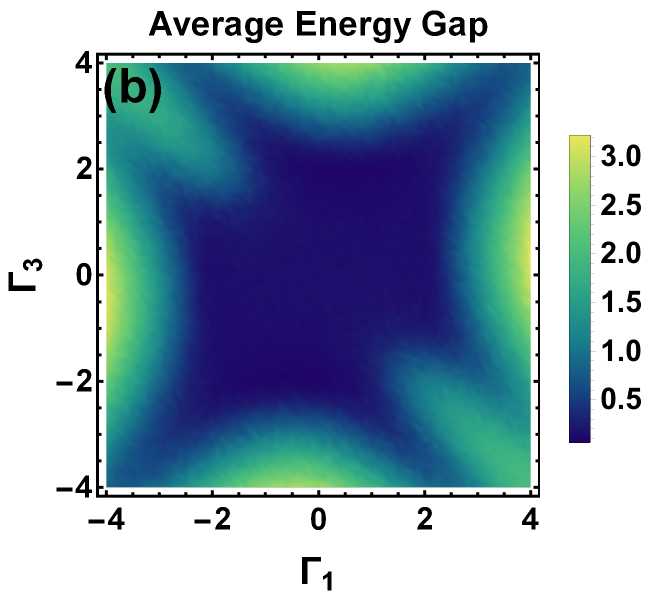}\hspace{0.5cm}
	\includegraphics[height=3.57cm]{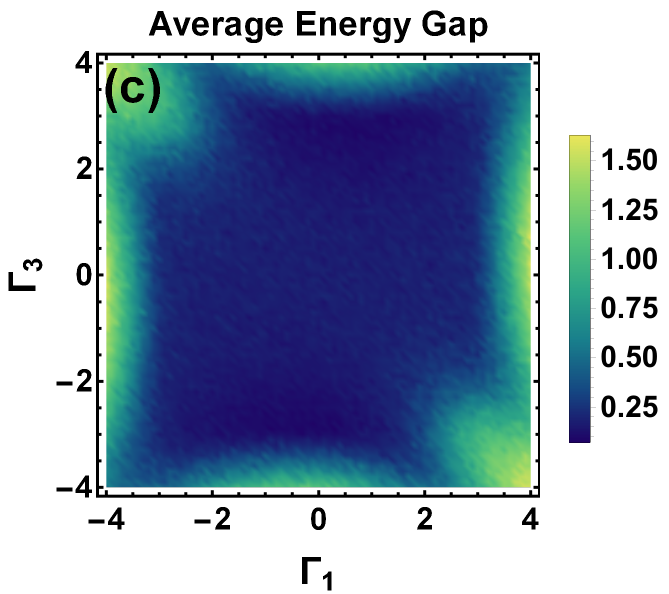}
\\
		\includegraphics[height=3.6cm]{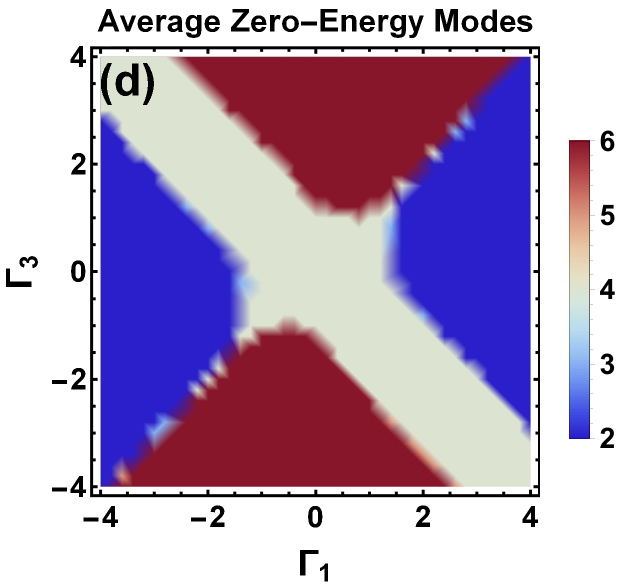}\hspace{0.5cm}
		\includegraphics[height=3.6cm]{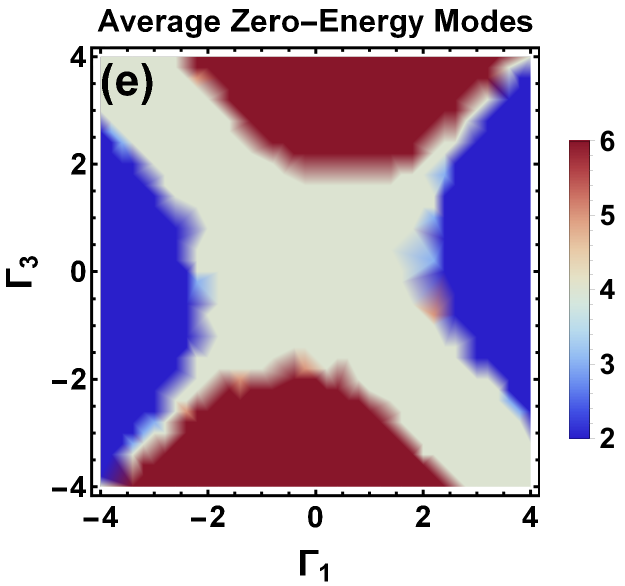}\hspace{0.5cm}
		\includegraphics[height=3.6cm]{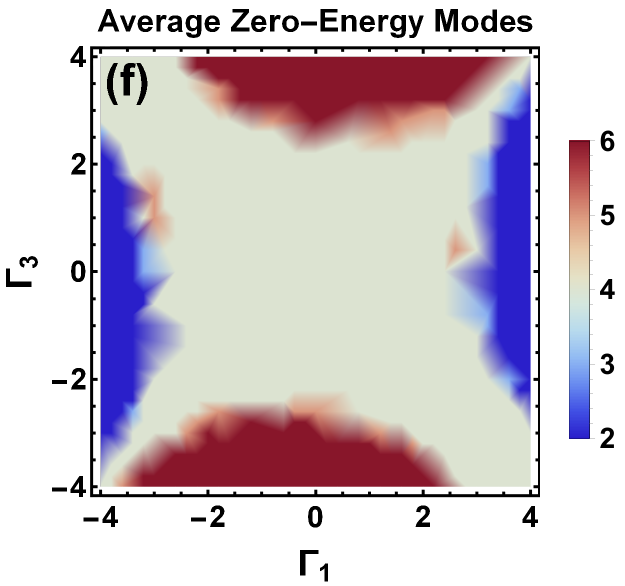}
\\
\includegraphics[height=2.9cm,width=4.0cm]{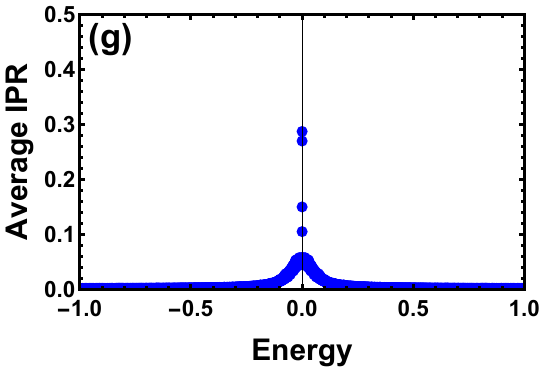}\hspace{0.5cm}
\includegraphics[height=2.9cm,width=4.0cm]{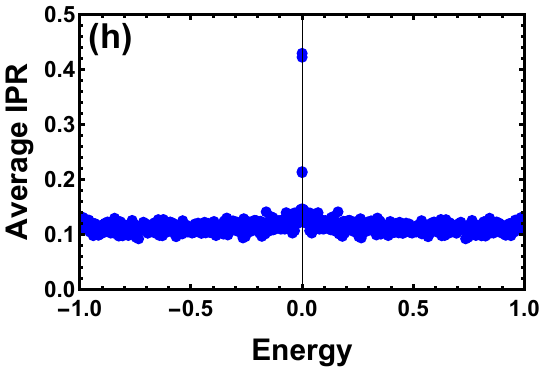}\hspace{0.5cm}
\includegraphics[height=2.9cm,width=4.0cm]{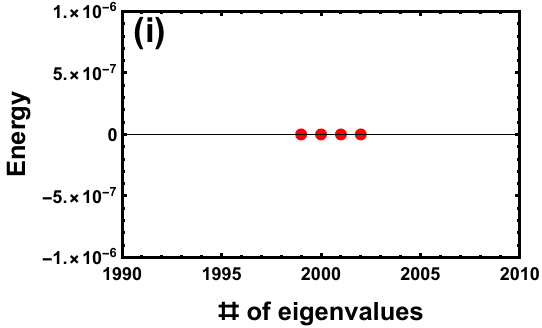}
\caption{Effect of disorder on the topological phase diagram for $\Gamma_0=0.2$, $\Gamma_2=1$ as a function of $\Gamma_1$ and $\Gamma_3$ for different disorder strengths. In the first row, the phase diagrams show the gap magnitude obtained for $N=60$, averaged over $50$ realizations, and with disorder strengths given by (a) $\eta_1=\eta_2=1$, (b) $\eta_1=\eta_2=2$, and (c) $\eta_1=\eta_2=3$. In the second row, phase diagrams show the average number of zero-energy eigenvalues for $N=450$ and $10$ realizations, again for disorder strengths given by (d) $\eta_1=\eta_2=1$, (e) $\eta_1=\eta_2=2$, and (f) $\eta_1=\eta_2=3$. Figures (g) and (h) shows the IPR for $\eta_1=\eta_2=0.1$ and $\eta_1=\eta_2=3$ respectively, at $(\Gamma_1,\Gamma_3)=(0.8,0.4)$ for $N=2000$ and averaged over $50$ realizations. Figure (i) shows the zero-energy eigenvalues at $(\Gamma_1,\Gamma_3)=(0.8,0.4)$ for $\eta_1=\eta_2=3$ and $N=2000$.
}
\label{PD_Gap1}
\end{figure*}

\section{Effects of disorder} \label{disorder}

In this section, we study the effect of disorder on the topological phase diagram and the multicritical points, $M_{1,2}$, of the model. We consider a lattice-dependent random disorder in the hoppings $\Gamma_1$ and $\Gamma_3$ such that,
\begin{align}
\tilde{\Gamma}_1(i)&= \Gamma_1+\gamma_1(i)\\
\tilde{\Gamma}_3(i)&= \Gamma_3+\gamma_2(i)
\end{align}
where $\gamma_{1,2}(i)$ take random values between $[-\eta_{1,2},\eta_{1,2}]$ for different disorder realizations.  
Figure~\ref{PD_Gap1} shows the phase diagram based on the energy gap for disorder strengths $\eta_1=\eta_2=1,2,\;\text{and}\;3$ in panels (a), (b), and (c), respectively. We observe the critical lines fading into a gapless region with increasing disorder strengths, tearing the gapped $w=2$ phase into two halves at $(\Gamma_1,\Gamma_3)=(0,0)$. This gapless phase dominates over the gapped phases at large disorder strengths. Compared to the clean case, we observe the multicritical points at $(\Gamma_1,\Gamma_3)=(\pm 0.8,\pm 0.4)$ vanish for strong disorder. To characterize the topological properties at these gapped and gapless phases, we scan the whole parameter space for the number of zero-energy modes. The panels (d), (e), and (f) in Fig.~\ref{PD_Gap1} show the phase diagrams based on the number of zero-energy eigenvalues for various disorder strengths, i.e., $\eta_1=\eta_2=1,2,\;\text{and}\;3$, respectively. 
We identify that the various gapped phases with the two, four, and six zero-energy eigenvalues, which are consistent with the $w=1,2 \;\text{and}\; 3$, respectively, obtained in the clean case. Moreover, we also identify the gapless phase with four zero-energy eigenvalues, which dominate the parameter space for higher disorder strengths. These phase diagrams indicate that the multicritical points $M_{1,2}$, where the gapped and gapless phases meet, persist in the presence of weak disorder. As the disorder strength increases, these points shift toward higher values of $|\Gamma_1|$ and $|\Gamma_3|$. Within our numerical resolution, we do not observe a sharp threshold disorder strength at which the multicritical points are destroyed; rather, their evolution with increasing disorder appears to be gradual.

To understand the localization properties at the gapless phase, we study the inverse participation ratio (IPR), defined as $\text{IPR}=\sum_j |\psi_j|^4$, where $\psi_j$ is the eigenstate at site $j$. IPR gives a direct measure of the localization strength of the eigenstate and identifies the presence of Anderson localization. While a zero IPR value indicates the extended states, non-zero values reflect spatial confinement of probability density. Figure.~\ref{PD_Gap1}(g) and (h) show the IPR for $\eta_1=\eta_2=0.1 \; \text{and}\; 3$ respectively, at $(\Gamma_1,\Gamma_3)=(0.8,0.4)$ which now a representative point in the gapless phase. We observe that the IPR for the bulk states (with non-zero energy) are almost zero for weak disorder; however, for higher disorders, the IPR clearly shows localized states. This indicates that the gapless phase is an Anderson-localized phase. The error bars and system size dependence of the IPR are discussed in Appendix.\ref{appc}. Moreover, IPR shows the signatures of zero-energy edge modes at the gapless phase with four eigenvalues with zero energy. These zero-energy modes are more localized than the bulk states with higher IPR values.
To further confirm the presence of zero-energy modes, we analyze the eigenvalue distribution at the same point in the parameter space for $\eta_1=\eta_2=3$, as shown in Figure~\ref{PD_Gap1}(i). We observe that two pairs of eigenvalues are at zero energy, representing a topologically nontrivial disordered chain with two edge modes at each end of the chain. These edge modes remain well localized even for higher disorder strengths. 

\begin{figure}[t]
	\includegraphics[width=4.1cm,height=2.9cm]{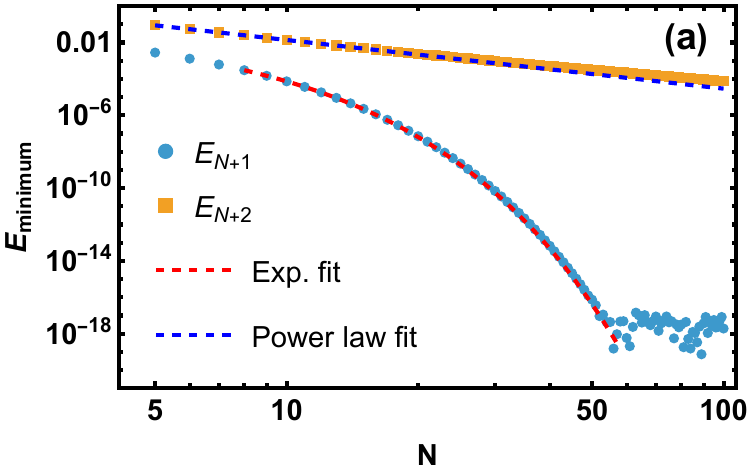}
    \hspace{0.2cm}
    \includegraphics[width=4.1cm,height=2.9cm]{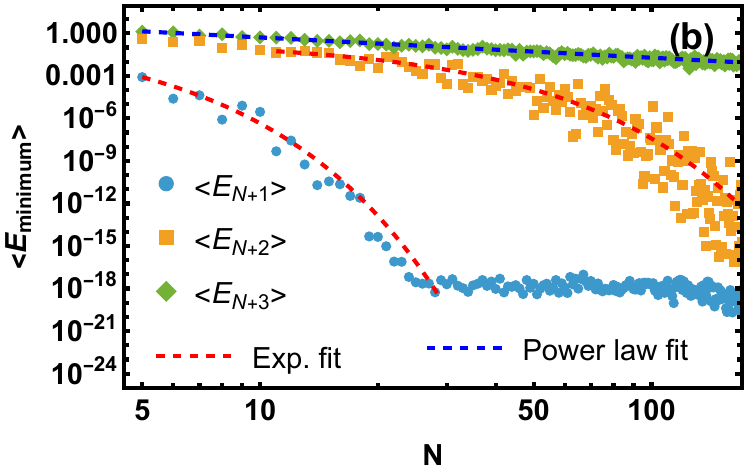}
	\caption{Finite-size scaling (plotted in log-log scale) of zero energy edge modes and bulk modes at the topological multicritical points. The plots are obtained for the parameters $\Gamma_{0}=0.2$, $\Gamma_{1}=\pm 0.8$, $\Gamma_{2}=1$ and $\Gamma_{3}=\pm 0.4$. (a) For a clean case $\eta_1=\eta_2=0$. We observe exponential scaling of $E_{N+1}$ and power-law scaling of $E_{N+2}$. (b) For $\eta_1=\eta_2=3$ and averaged over $10$ disorder realizations. Exponential scaling is obtained for $\left\langle E_{N+1}\right\rangle$, $\left\langle E_{N+2}\right\rangle$ and power-law scaling is obatined for $\left\langle E_{N+3}\right\rangle$.
	}
	\label{fz}
\end{figure}

To establish the robustness of the zero-energy edge modes and to distinguish them from bulk states, we analyze the finite-size scaling behavior at the multicritical points in both clean and disordered systems. In the clean case, we focus on the smallest positive eigenvalue, namely the $(N+1)$th eigenvalue $E_{N+1}$, which corresponds to an edge mode since the spectrum consists of $2N$ eigenvalues for a system of size $N$. We also examine the next eigenvalue, $E_{N+2}$, which represents the lowest bulk excitation under open boundary conditions. The system-size dependence of these eigenvalues is shown in Fig.~\ref{fz}(a). We find that the edge-mode energy $E_{N+1}$ exhibits an exponential decay with system size, accompanied by a curvature, eventually reaching numerical noise at $E_{N+1}\approx 10^{-18}$. 
In contrast, the bulk-mode energy $E_{N+2}$ follows a power-law scaling with system size, appearing as a linear trend in the log–log plot.

We perform a similar analysis in the presence of disorder with strength $\eta_1=\eta_2=3$. In this case, we extract $E_{N+1}$, $E_{N+2}$, and $E_{N+3}$ and average them over many disorder realizations. The resulting finite-size scaling is presented in Fig.~\ref{fz}(b). We observe that both $\langle E_{N+1}\rangle$ and $\langle E_{N+2}\rangle$ decay exponentially with system size, indicating the persistence of zero-energy edge modes in the disordered regime. By contrast, $\langle E_{N+3}\rangle$ exhibits a power-law dependence on system size, consistent with its bulk character.

So far, throughout the analysis, the disorder strengths are kept equal, i.e., $\eta_1=\eta_2$. However, a more general case where $\eta_1\neq\eta_2$ can also be studied, which yields qualitatively similar results. The details for two cases with $\eta_1<\eta_2$ and $\eta_1>\eta_2$ are discussed in Appendix.~\ref{app3}. Overall, we observe that for small values of the disorder, the multicritical points do not vanish, but their position in the parameter space is merely shifted.
For higher values of the disorder, all the topological gaps eventually close, and all multicritical points and critical lines disappear, realizing a transition to a gapless Anderson localized phase.
Moreover, the Anderson localized phase is topologically nontrivial and characterized by the presence of two zero-energy modes localized at each end of the chain. 

\section{Conclusion}
In this work, we demonstrated that in a topological chain with extended nearest neighbors, topological characters can also be associated with the multicritical points apart from quantum critical points under certain conditions on the parameter space. In a generic model that encapsulates 1D topological insulators and superconductors, at first, we calculated the winding number and identified all the topologically gapped phases. We obtained a phase diagram containing a plane with only topologically nontrivial phases $w=1,2,3$. Besides, the quantum critical phases of the model with their topological nature have been characterized using topological invariants calculated from the zeros of a complex function associated with the Hamiltonian. There are $w_c=1,2$ phases on the high symmetry critical lines and only $w_c=1$ on the non-high symmetry critical lines. These distinct quantum critical phases meet at a multicritical point with quadratic dispersion. As the phase diagram is completely nontrivial with both gapped and critical nontrivial phases, it produces topologically nontrivial multicritical points characterized by $w_{mc}=1$. The outcomes have been verified numerically under open boundary conditions using eigenvalue and probability distributions. 

In order to provide an alternative way to look at the phase diagram and the multicritical lines, we analyze the nature of zeros or roots around the unit circle on a complex plane in the vicinity of the multicritical points. Among the three roots of the cubic polynomial associated with the Hamiltonian, we identify one of the roots that lies inside the unit circle, causing $w_{mc}=1$, i.e., the localization of one stable edge mode at each end of the open chain. The corresponding localization length, eigenvalue, and probability distributions agree with the topological invariant. The quadratic nature of multicriticality is manifested as degenerate roots on the unit circle, leading to 
the dynamical critical exponent $z=m=2$ that has been computed separately for both multicritical points. The mobility of the roots near the multicritical point along various critical phases indicates that complex conjugate roots confluence on the real axis, yielding degenerate zeros at $\pm1$ and a zero inside the unit circle, finally leading to $w_{mc}=1$ at the multicritical points. 

Further, we calculated the discriminant of the cubic polynomial and 
characterized its unique behavior around the topological multicritical point. Uniquely, the discriminant $\Delta=0$ at the multicritical points and flips the sign across these points on the high-symmetry critical lines. Such behavior is absent at the multicritical points in different parameter spaces, which supports trivial phases ($w=w_c=0$). Therefore, we have argued that the condition for the parameter space to satisfy to get topological multicritical points can be obtained as $\Gamma_{0}< \Gamma_{2}/3$. Under this condition, the parameter space is completely nontrivial with all topologically gapped and critical phases, including multicriticalities. This picture is further supported by identifying a kinetic inversion mechanism in the higher-order terms, which facilitates the transition from trivial to non-trivial multicritical points at $\Gamma_{0}=\Gamma_{2}/3$.

The multicritical points analyzed here are robust against weak disorder, although they vanish at stronger disorder strength. Interestingly, strong disorder drives the system into a gapless and topologically nontrivial Anderson-localized phase characterized by the presence of two zero-energy modes localized at the end of the chain.

The nontrivial multicritical points obtained in this work host one edge mode characterized by the topological invariants $w_{mc}=1$. However, in general, multicritical points with $w_{mc}>1$ can also be realized by obtaining higher winding number gapped and critical phases in the phase diagram. This can be achieved by increasing the nearest neighbor couplings in the 1D topological chains. 
These setups can be realized in various platforms, including, for instance,  superconducting circuits \cite{PhysRevB.101.035109,niu2021simulation}, ultracold atoms with a good control over nearest neighbor couplings \cite{goldman2016topological,meier2016observation}, quantum walks \cite{kumar2025topological,PhysRevB.88.121406,PhysRevLett.118.130501}. 
These systems can exhibit non-local or long-range interactions, allowing particles to tunnel between distant sites.
Moreover, note that the fundamental ingredient for the existence of topologically nontrivial multicritical points is the presence of an intersection of critical lines separating topological phases that are all nontrivial.
Physically, the existence of such multicritical points is made possible by the presence of next-to-nearest neighbor hoppings.
This can be realized in principle in any topological symmetry class, and it is not restricted to the spinless SSH and Kitaev models considered here.
Therefore, we believe that the topologically nontrivial multicritical points introduced here can be experimentally realized and observed in realistic physical systems with long-range non-local interactions. 
Moreover, we can expect that the unique characteristics of multicritical points will be exploited in technological applications in the field of topological quantum computation, where topologically protected localized edge modes carry the quantum information \cite{10.21468/SciPostPhys.3.3.021}. 
Our model can provide 
a regime with a fully nontrivial parameter space with topological critical lines and multicritical points.
In particular, on these points, edge modes do not delocalize into the bulk due to topological transitions. 
This makes the corresponding quantum information protected without any decoherence arising due to the phase transitions.

\begin{acknowledgements}
	R.~R.~K. acknowledges the Indian Institute of Technology (IIT), Bombay for support through the Institution Post-Doc program.
	P.~M. is partially supported by the Japan Society for the Promotion of Science (JSPS) Grant-in-Aid for Early-Career Scientists KAKENHI Grant~No.~23K13028 and No.~20K14375, and Grant-in-Aid for Transformative Research Areas (A) KAKENHI Grant~No.~22H05111.
\end{acknowledgements}

\appendix
\section{Edge modes at quantum critical phases}\label{ED-Crit}
\begin{figure}[t]
	\includegraphics[width=4.2cm,height=2.8cm]{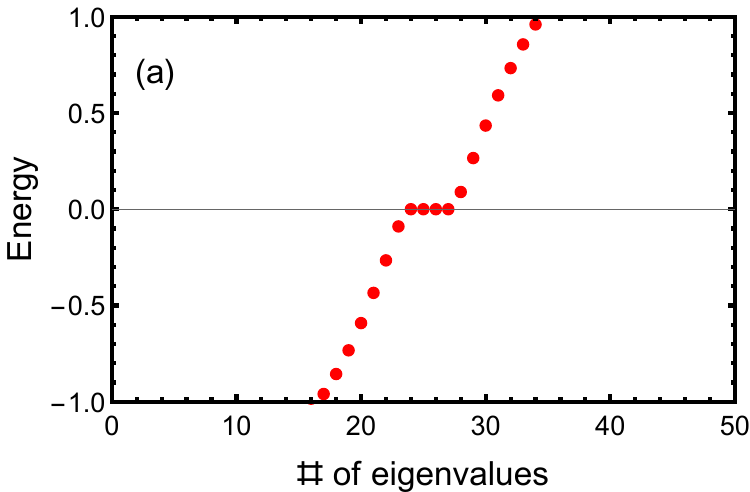}\hspace{0.2cm}\includegraphics[width=4.2cm,height=2.8cm]{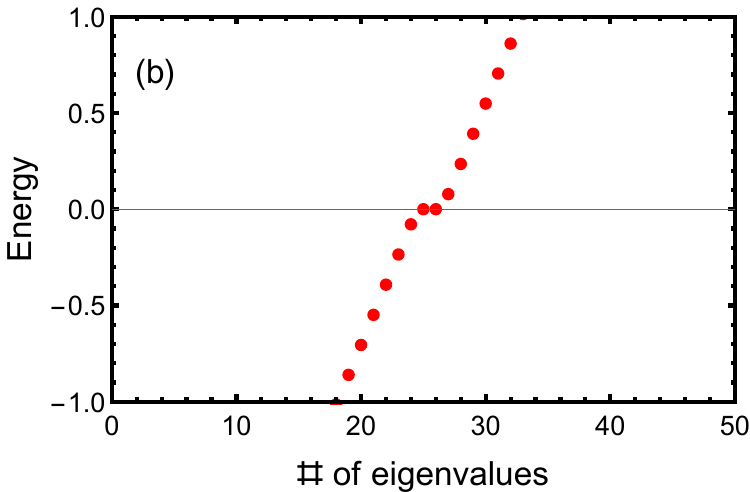}\\
	\includegraphics[width=4.2cm,height=2.8cm]{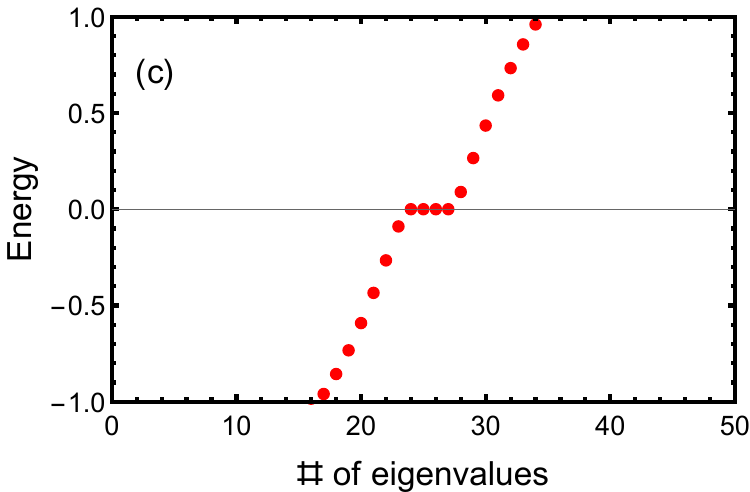}\hspace{0.2cm}\includegraphics[width=4.2cm,height=2.8cm]{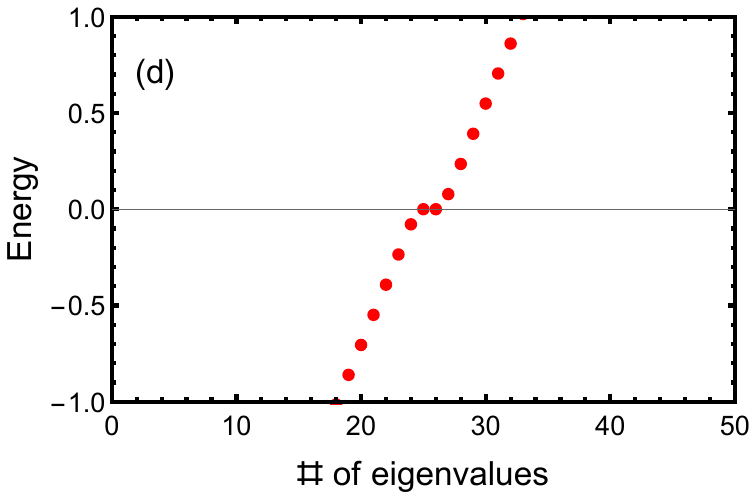}\\
	\includegraphics[width=4.2cm,height=2.8cm]{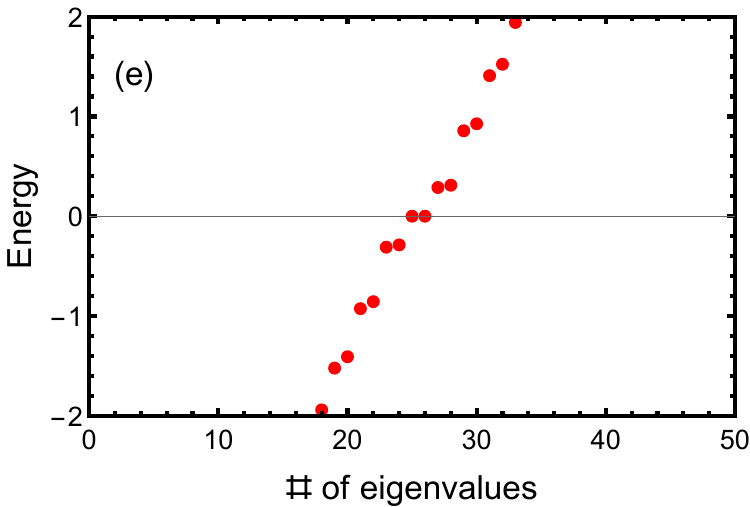}\hspace{0.2cm}\includegraphics[width=4.2cm,height=2.8cm]{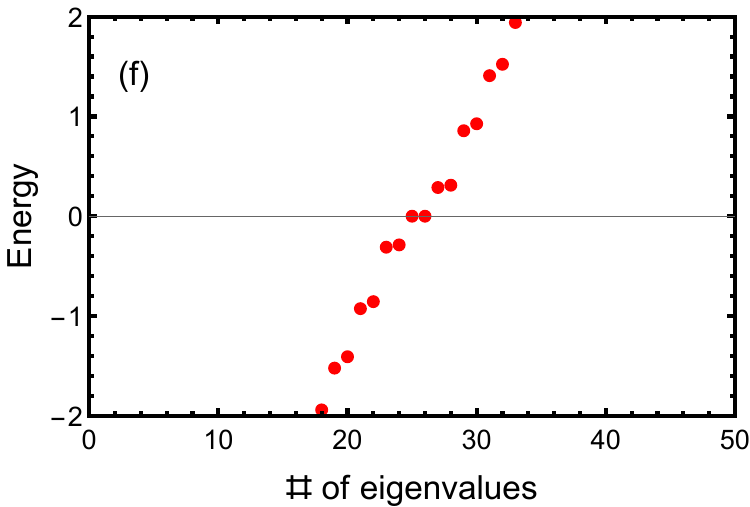}\\
	\caption{Eigenvalue distribution under open boundary condition. The eigenvalues near the zero energy are shown for the parameters $\Gamma_0=0.2$ and $\Gamma_2=1$. Critical phases between (a) $w=2$ and $w=3$ [at $(\Gamma_1,\Gamma_3)=(0.1,1.1)$] and (b) $w=2$ and $w=1$ [at $(\Gamma_1,\Gamma_3)=(1.4,-0.2)$] on the blue critical line, show two pairs and a single pair of eigenvalues that are localized at the zero energy, respectively. Critical phases between (c) $w=2$ and $w=3$ [at $(\Gamma_1,\Gamma_3)=(-0.1,-1.1)$] and (d) $w=2$ and $w=1$ [at $(\Gamma_1,\Gamma_3)=(-1.4,0.2)$] on the red critical line, show two pairs and a single pair of eigenvalues that are localized at the zero energy respectively. Critical phases between $w=3$ and $w=1$ (e) on the magenta critical line [at $(\Gamma_1,\Gamma_3)=(2.5,2.43)$] and (f) on the green critical line [at $(\Gamma_1,\Gamma_3)=(-2.5,-2.43)$], show a single pair of eigenvalues are localized at the zero energy respectively.}
	\label{ED}
\end{figure}
 
On the blue critical line we find the critical phase between $w=2$ and $w=3$ hosts two edge modes as shown in Fig.~\ref{ED}(a). The eigenvalue distribution shows that the bulk is gapless and there are two pairs of eigenvalues at zero energy, which correspond to the two edge modes localized at each end of an open chain. At the critical phase between $w=2$ and $w=1$, we find only a pair of eigenvalues, as shown in Fig.~\ref{ED}(b). The transition between these distinct critical phases on the critical line occurs at the multicritical point $M_1$. Similarly, we observe that on the red critical line, the critical phase between $w=2$ and $w=3$ possesses two whilst the critical phase between $w=2$ and $w=1$ possesses one edge mode, as shown in Fig.~\ref{ED}(c) and Fig.~\ref{ED}(d) respectively. These two distinct critical phases are separated by the multicritical point $M_2$. 
Non-high-symmetry critical lines host only one kind of critical phase, which meets the high-symmetry phases at the multicritical points. The eigenvalue distribution for both the critical phases, on magenta and green lines between $w=3$ and $w=1$, shows one pair of eigenvalues indicating one edge mode at each end of the chain, as shown in Fig.~\ref{ED}(e) and Fig.~\ref{ED}(f). Topological transitions between high-symmetry critical phases and non-high-symmetry critical phases also occur at the multicritical points $M_{1,2}$.

\section{Topological invariants at quantum critical phases}\label{W-crit}
\begin{figure}[t]
	\includegraphics[width=3.3cm,height=3.0cm]{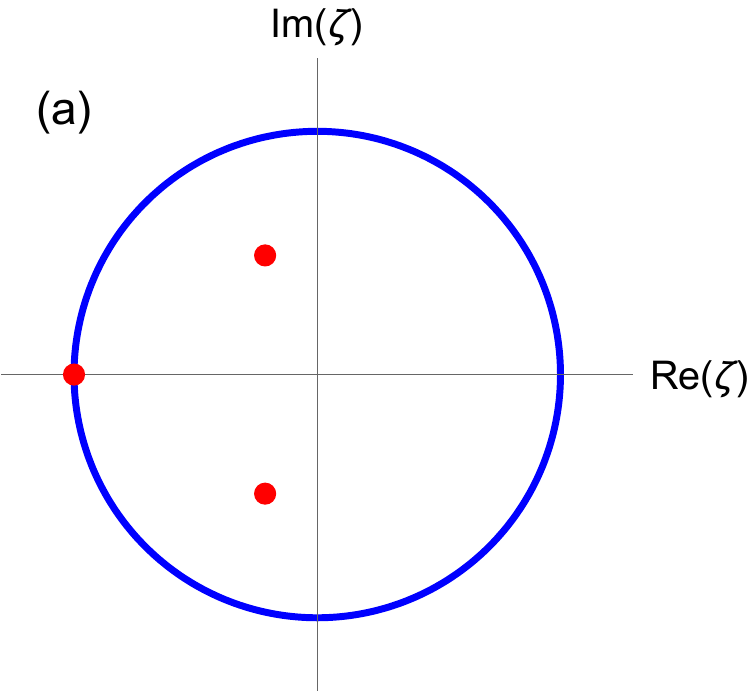}\hspace{0.2cm}\includegraphics[width=3.3cm,height=3.0cm]{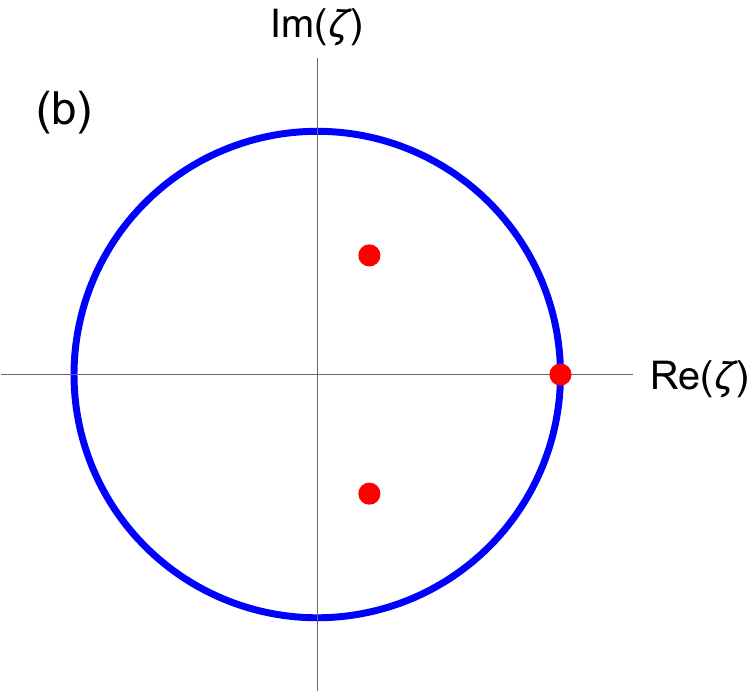}\\
	\includegraphics[width=3.3cm,height=3.0cm]{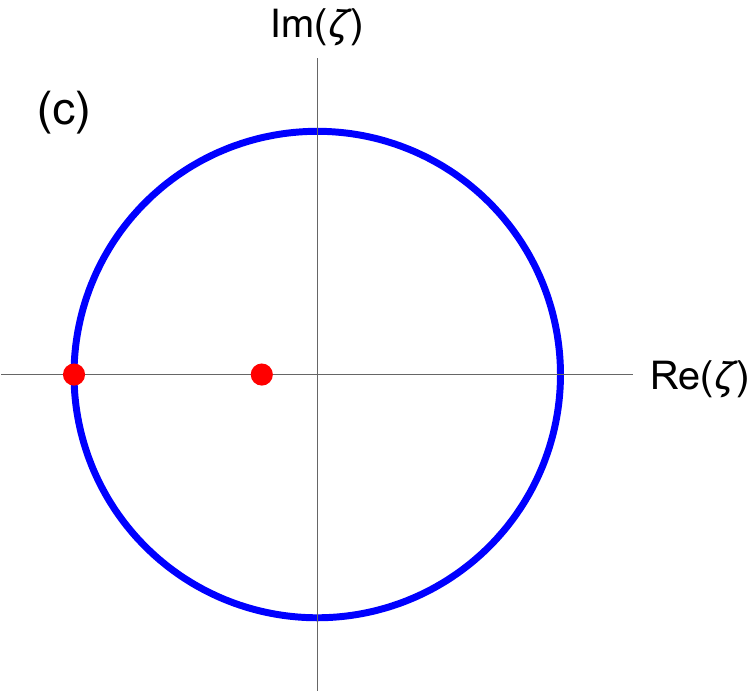}\hspace{0.2cm}\includegraphics[width=3.3cm,height=3.0cm]{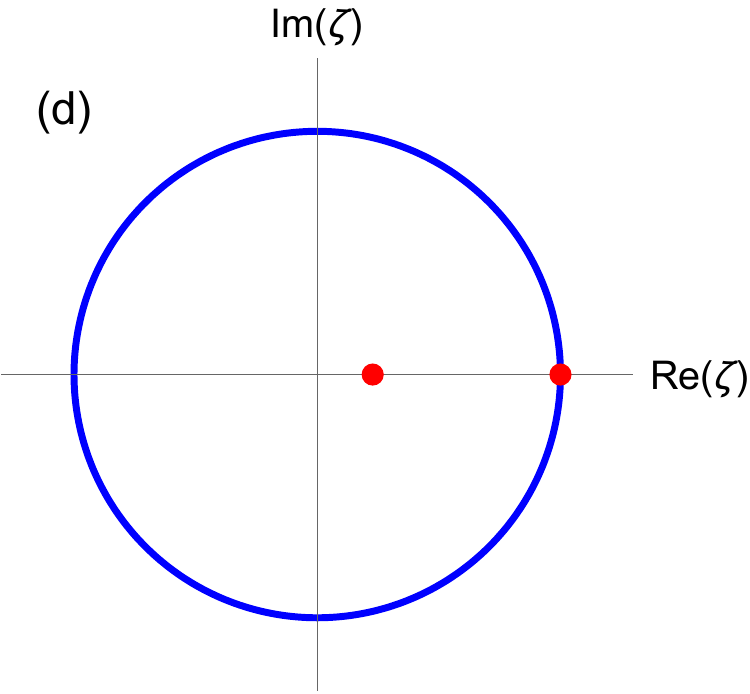}\\
	\includegraphics[width=3.3cm,height=3.0cm]{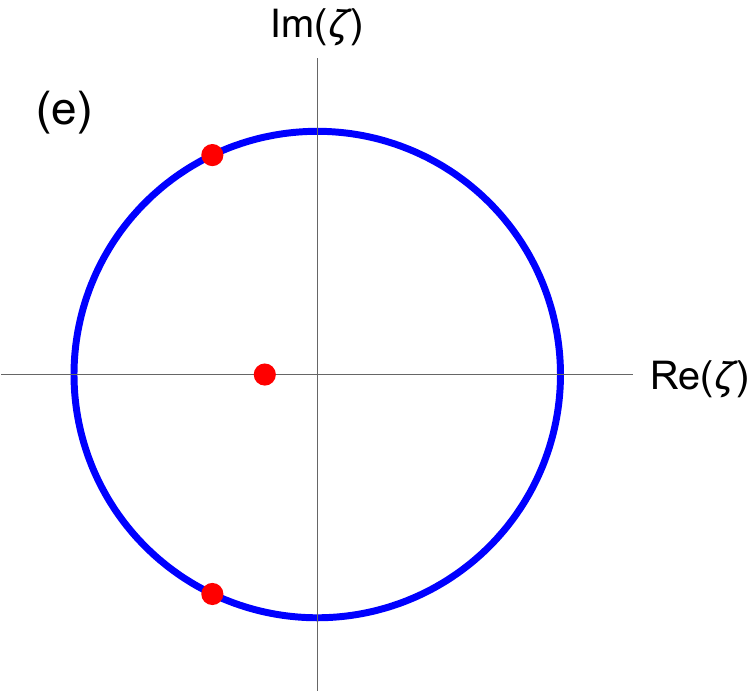}\hspace{0.2cm}\includegraphics[width=3.3cm,height=3.0cm]{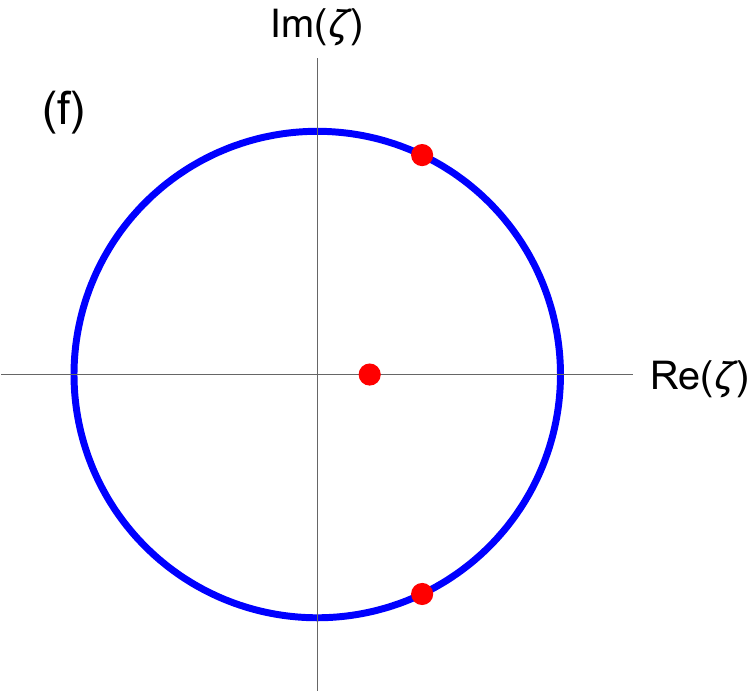}\\
	\caption{Zeros of the complex function associated with the Hamiltonian on a complex plane with unit circle. The zeros are given in Eqs.\ref{zero1}, \ref{zero2}, and \ref{zero3}. Critical phases between $w=2$ and $w=3$ (a) on the blue critical line [at $(\Gamma_1,\Gamma_3)=(0.1,1.1)$] and (b) on the red critical line [at $(\Gamma_1,\Gamma_3)=(-0.1,-1.1)$], show two zeros inside and one zero on the unit circle, leading to $w_c=2$ and $c=1/2$. Critical phases between $w=2$ and $w=1$ (c) on the blue critical line [at $(\Gamma_1,\Gamma_3)=(1.4,-0.2)$] and (d) on the red critical line [at $(\Gamma_1,\Gamma_3)=(-1.4,0.2)$], show one zero inside and one zero on the unit circle, leading to $w_c=1$ and $c=1/2$. Critical phases between $w=3$ and $w=1$ (e) on the magenta critical line [at $(\Gamma_1,\Gamma_3)=(2.5,2.43)$] and (f) on the green critical line [at $(\Gamma_1,\Gamma_3)=(-2.5,-2.43)$], show one zero inside and two zeros on the unit circle, leading to $w_c=1$ and $c=1$.}
	\label{ZC}
\end{figure}
In Fig.~\ref{ZC}, the zeros at the various critical phases of the model is shown. At the critical phases between $w=2$ and $w=3$, we find two zeros that lie inside the unit circle, leading to $w_c=2$, as shown in Fig.~\ref{ZC}(a) and (b). One of the zeros lies on the unit circle at $\pm1$ leading to $c=1/2$. Therefore, this critical phase hosts two edge modes localized at each end of the open chain with critical properties described by $c=1/2$. At the critical phases between $w=2$ and $w=1$, we find only one zero that lies inside and one that lies on the unit circle, leading to $w_c=1$ and $c=1/2$ respectively, as shown in Fig.~\ref{ZC}(c) and (d). Similarly, at the critical phases between $w=3$ and $w=1$, there is one zero that lie inside and two zeros lie on the unit circle, leading to the same $w_c=1$ but different $c=1$ phase, as shown in Fig.~\ref{ZC}(e) and (f). Therefore, the value of $c$ identifies the high symmetry and non-high symmetry criticalities. The value of $w_c$ obtained for various critical phases is consistent with the eigenvalue distribution under open boundary conditions shown in Fig.~\ref{ED}. These results show that all the critical phases of the model are topologically nontrivial phases.

\section{Standard deviation and fractal dimension for the IPR calculations}\label{appc}
\begin{figure}[t]
\includegraphics[width=4.1cm,height=3.5cm]{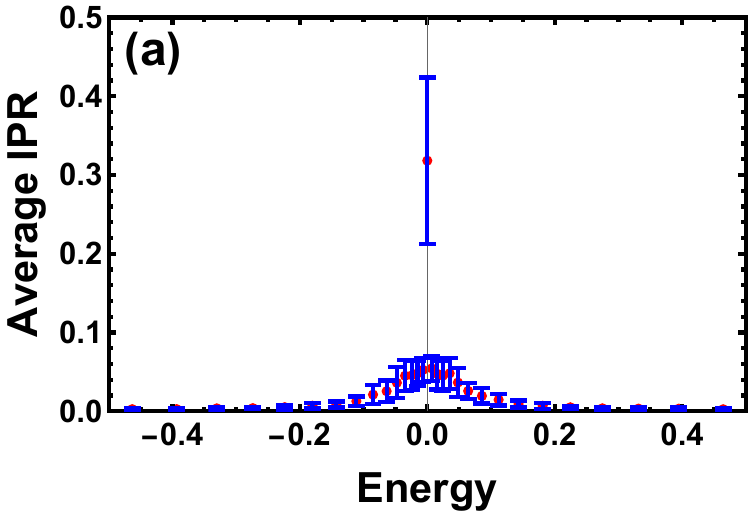} \hspace{0.1cm}
\includegraphics[width=4.1cm,height=3.5cm]{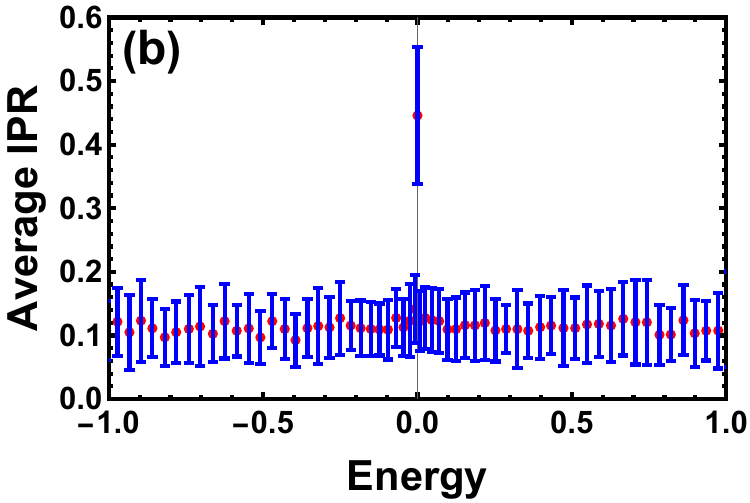}\\ \vspace{0.5cm}
\includegraphics[height=3.9cm]{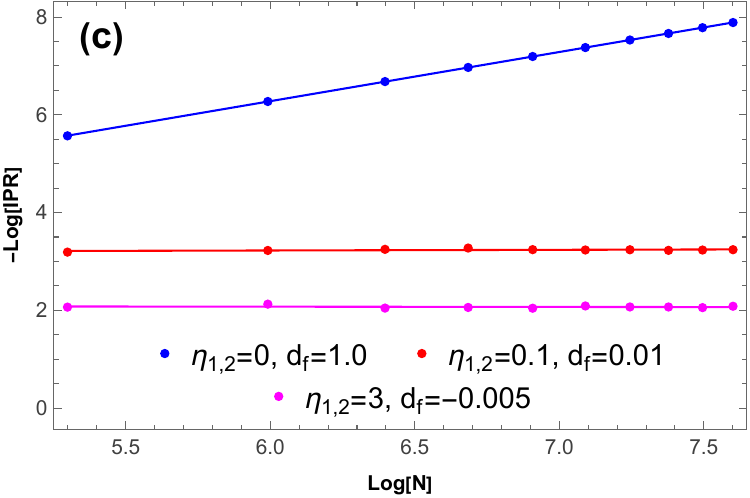}
\caption{Error bars based on the standard deviation $\sigma$, for IPR calculations are shown in (a) for $\eta_1=\eta_2=0.1$ and (b) for $\eta_1=\eta_2=3$. Other parameters are kept same as in Fig.\ref{PD_Gap1}(g,h) i.e., $\Gamma_0=0.2$, $\Gamma_1=0.8$, $\Gamma_2=1$, and $\Gamma_3=0.4$. (c) System size scaling of IPR values at the multicritical points with $\Gamma_0=0.2$, $\Gamma_1=\pm 0.8$, $\Gamma_2=1$ and $\Gamma_3=\pm 0.4$. The bulk states close to zero energy i.e., within the energy window $0.01>E>10^{-4}$ are considered. For clean case ($\eta_1=\eta_2=0$) we get fractal dimension $d_f=1$ representing an extended phase. For weak ($\eta_1=\eta_2=0.1$) and strong ($\eta_1=\eta_2=3$) disorders, we get $d_f=0.01$ and $d_f=-0.005$ respectively representing Anderson localized phases.}
\label{fd}
\end{figure}
To show the error bars of the average IPR calculated in Fig.~\ref{PD_Gap1}(g) and (h), we compute the standard deviation $\sigma$ over the same parameter space with system size $N=2000$ and disorder realizations $50$. The standard deviation can be obtained as $\sigma =\sqrt{\langle\text{IPR}^2\rangle - \langle\text{IPR}\rangle^2}$.
Fig.~\ref{fd}(a) and (b) present the energy-resolved average IPR along with vertical error bars for $\eta_1=\eta_2=0.1$ and $\eta_1=\eta_2=3$, respectively. For clarity, only a subset of the energy points is displayed. Across the entire energy spectrum, we find that $\sigma<\langle\text{IPR}\rangle$, indicating relatively weak sample-to-sample fluctuations and suggesting that the disorder-averaged IPR is statistically stable. A modest enhancement of $\sigma$ is observed near zero energy, implying slightly stronger fluctuations in this region. This behavior may reflect enhanced sensitivity of the corresponding eigenstates to disorder near the band center.

The system-size dependence of the IPR follows the scaling relation $\text{IPR} \propto N^{-d_f}$, where $N$ denotes the system size and $d_f$ is the fractal dimension \cite{lu2022exact,SIRCAR2026131192}. For an extended phase, one expects $d_f=1$, whereas for a localized phase $d_f=0$. Taking the logarithm of the above scaling relation, the fractal dimension can be extracted as
\begin{equation}
    d_f = -\frac{\log(\mathrm{IPR})}{\log(N)} + c
\end{equation}
where $c$ is a constant offset arising from finite-size effects. Consequently, $d_f$ is given by the slope of the linear fit in a $\log(\mathrm{IPR})$ versus $\log(N)$ plot.

We now present the fractal dimension extracted at the multicritical points for both clean and disordered systems. Since the multicritical points host gapless bulk excitations, we focus on the IPR of bulk states close to zero energy and analyze their scaling with system size. Fig.~\ref{fd}(c) shows the system-size scaling of the IPR together with the corresponding values of $d_f$ obtained from numerical fits. In the clean limit, we find $d_f=1$, indicating that the bulk states are extended. Upon introducing weak disorder with $\eta_1=\eta_2=0.1$, the IPR becomes nearly independent of system size, yielding $d_f\simeq 0.01$, which signals the onset of localization. For stronger disorder, $\eta_1=\eta_2=3.0$, we obtain $d_f\simeq -0.005$, consistent with an Anderson-localized phase.

\section{Effects of unequal disorder strengths}\label{app3}
\begin{figure}[t]
	\includegraphics[height=3.9cm]{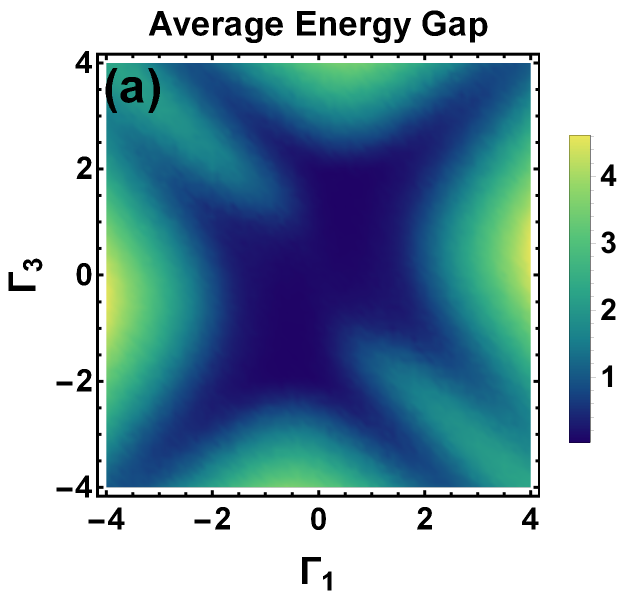}
	\includegraphics[height=3.9cm]{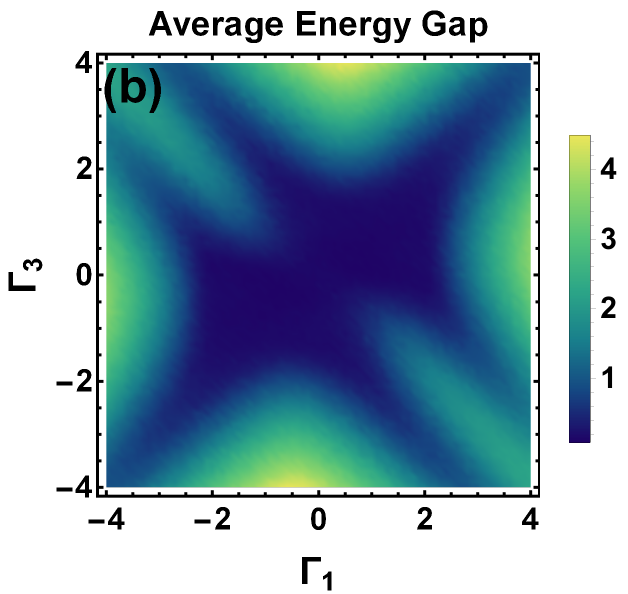}
\\
		\includegraphics[height=3.9cm]{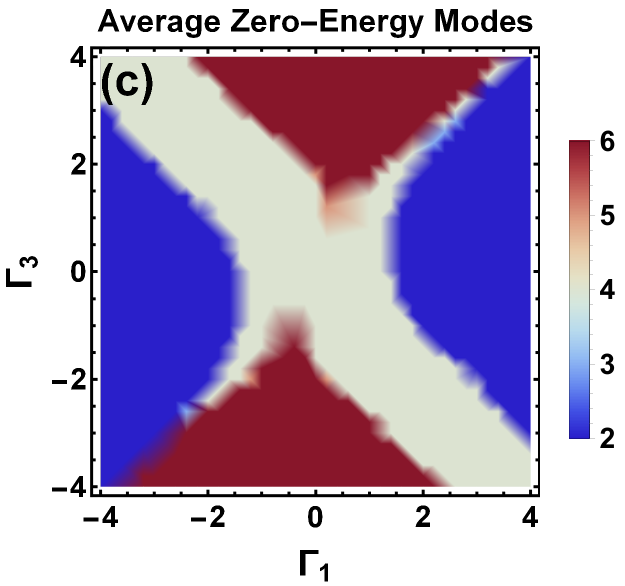}
		\includegraphics[height=3.9cm]{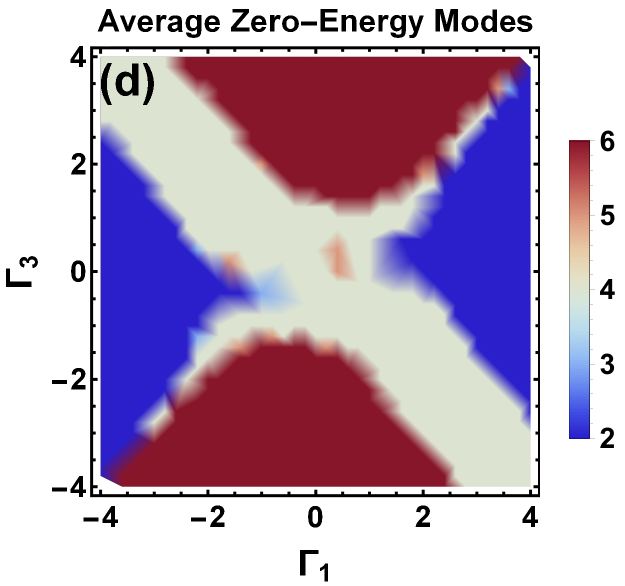}
\caption{Phase diagrams for unequal disorder strength with $\Gamma_0=0.2$, $\Gamma_2=1$. The first row shows the gap magnitude obtained for $N=60$, averaged over $50$ realizations. The second row shows the average number of zero-energy eigenvalues for $N=450$ and $10$ realizations. 
The disorder strengths in (a) and (c) is $\eta_1<\eta_2:\;\eta_1=0.5, \;\eta_2=2$. Similarly for (b) and (d) is $\eta_1>\eta_2:\;\eta_1=2, \;\eta_2=0.5$.}
\label{pdueq}
\end{figure}
To understand the effects of disorder with unequal strengths, we consider the two cases: (i) $\eta_1<\eta_2$ and (ii) $\eta_1>\eta_2$. Due to the unequal disorder strengths, as shown in Fig.\ref{pdueq}, the phase diagram breaks the symmetric nature along $\Gamma_3 = \Gamma_1$ with a shift depending on the interaction strengths. However, other qualitative features discussed in Section.~\ref{disorder} remain the same. 
As shown in Fig.~\ref{pdueq}(a) and (b), the phase diagram based on the energy gap shows the Anderson localized phase, which dominates for higher disorder strengths for both cases. The topological properties of these phases identified from the average number of zero energy eigenvalues, as shown in the phase diagrams in Fig.~\ref{pdueq}(c) and (d), reveal that the gapless phase hosts two pairs of zero energy edge modes for both cases. The topological multicritical points, while only shifting their position in the parameter space for weak disorders, vanish for higher disorder strengths, similar to the case $\eta_1=\eta_2$ discussed in Section.~\ref{disorder}.


\begin{thebibliography}{10}
	\bibitem{verresen2018topology}
Ruben Verresen, Nick~G Jones, and Frank Pollmann.
	\newblock Topology and edge modes in quantum critical chains.
\href{https://journals.aps.org/prl/abstract/10.1103/PhysRevLett.120.057001}{\newblock {\em Phys. Rev. Lett.} \textbf{120}, 057001 (2018).}

\bibitem{verresen2019gapless}
Ruben Verresen, Ryan Thorngren, Nick~G Jones, and Frank Pollmann.
	\newblock Gapless topological phases and symmetry-enriched quantum criticality.
\href{https://journals.aps.org/prx/abstract/10.1103/PhysRevX.11.041059}{\newblock {\em Phys. Rev. X.} \textbf{11}, 041059 (2021).} 

\bibitem{jones2019asymptotic}
Nick~G Jones and Ruben Verresen.
\newblock Asymptotic correlations in gapped and critical topological phases of 1d quantum systems.
\href{https://link.springer.com/article/10.1007/s10955-019-02257-9}{\newblock {\em J. Stat. Phys.} \textbf{175}, 1164--1213 (2019).}

\bibitem{verresen2020topology}
Ruben Verresen.
	\newblock Topology and edge states survive quantum criticality between topological insulators.
\href{https://arxiv.org/abs/2003.05453}{\newblock {\em  arXiv:2003.05453v1 [cond-mat.str-el]}} (2020).

\bibitem{rahul2021majorana}
S~Rahul, Ranjith~R Kumar, Y R~Kartik, and Sujit Sarkar.
	\newblock Majorana zero modes and bulk-boundary correspondence at quantum criticality.
\href{https://journals.jps.jp/doi/abs/10.7566/JPSJ.90.094706?journalCode=jpsj}{\newblock {\em J. Phys. Soc. Jpn.} \textbf{90}, 094706 (2021).}

\bibitem{niu2021emergent}
Sen Niu, Yucheng Wang, and Xiong-Jun Liu.
	\newblock Emergent gapless topological luttinger liquid.
\href{https://arxiv.org/abs/2106.13400}{\newblock {\em arXiv:2106.13400v2 [cond-mat.str-el]}} (2021).


\bibitem{PhysRevB.104.075132}
Ryan Thorngren, Ashvin Vishwanath, and Ruben Verresen.
	\newblock Intrinsically gapless topological phases.
\href{https://journals.aps.org/prb/abstract/10.1103/PhysRevB.104.075132}{\newblock {\em Phys. Rev. B.} \textbf{104}, 075132 (2021).}

\bibitem{PhysRevResearch.3.043048}
Oleksandr Balabanov, Daniel Erkensten, and Henrik Johannesson.
	\newblock Topology of critical chiral phases: Multiband insulators and superconductors.
\href{https://journals.aps.org/prresearch/abstract/10.1103/PhysRevResearch.3.043048}{\newblock {\em Phys. Rev. Res.} \textbf{3}, 043048 (2021).}

\bibitem{fraxanet2021topological}
Joana Fraxanet, Daniel Gonz{\'a}lez-Cuadra, Tilman Pfau, Maciej Lewenstein, Tim
Langen, and Luca Barbiero.
	\newblock Topological quantum critical points in the extended bose-hubbard model.
\href{https://journals.aps.org/prl/abstract/10.1103/PhysRevLett.128.043402}{\newblock {\em Phys. Rev. Lett.} \textbf{128}, 043402 (2022).}

\bibitem{keselman2015gapless}
Anna Keselman, Erez Berg.
	Gapless symmetry-protected topological phase of fermions in one dimension.
\href{https://journals.aps.org/prb/abstract/10.1103/PhysRevB.91.235309}{\newblock {\em Phys. Rev. B.} \textbf{91}, 235309 (2015).}

\bibitem{scaffidi2017gapless}
Thomas Scaffidi, Daniel E. Parker, Romain Vasseur.
	Gapless symmetry-protected topological order.
\href{https://journals.aps.org/prx/abstract/10.1103/PhysRevX.7.041048}{\newblock {\em Phys. Rev. X.} \textbf{7}, 041048 (2017).}

\bibitem{duque2021topological}
Carlos M. Duque, Hong-Ye Hu, Yi-Zhuang You, Vedika Khemani, Ruben Verresen, and Romain Vasseur.
	Topological and symmetry-enriched random quantum critical points.
\href{https://journals.aps.org/prb/abstract/10.1103/PhysRevB.103.L100207}{\newblock {\em Phys. Rev. B.} \textbf{103}, L100207 (2021).}

\bibitem{kumar2021multi}
Ranjith~R Kumar, Y R~Kartik, S~Rahul, and Sujit Sarkar.
	\newblock Multi-critical topological transition at quantum criticality.
\href{https://www.nature.com/articles/s41598-020-80337-7}{\newblock {\em Sci. Rep.} \textbf{11}, 1--20 (2021).}

\bibitem{kwwangSciPostPhys.12.4.134}
Ke Wang, T. A. Sedrakyan.
	\newblock Universal finite-size amplitude and anomalous entanglement entropy of z=2 quantum Lifshitz criticalities in topological chains.
\href{https://www.scipost.org/SciPostPhys.12.4.134?acad_field_slug=chemistry}{\newblock {\em SciPost Phys.} \textbf{12}, 134 (2022).}

\bibitem{kumar2021topological}
Ranjith R. Kumar, Nilanjan Roy, Y. R. Kartik, S. Rahul, and Sujit Sarkar.
	\newblock Signatures of topological phase transition on a quantum critical line.
\href{https://journals.aps.org/prb/abstract/10.1103/PhysRevB.107.205114}{\newblock {\em Phys. Rev. B.}, \textbf{107} 205114 (2023).}

\bibitem{kumar2023topological}
Ranjith R. Kumar, Y. R. Kartik, and Sujit Sarkar.
	\newblock Topological phase transition between non-high symmetry critical phases and curvature function renormalization group.
\href{https://iopscience.iop.org/article/10.1088/1367-2630/aced1a}{\newblock {\em New J. Phys.}, \textbf{25} 083027 (2023).}

\bibitem{PhysRevLett.129.210601}
Xue-Jia Yu, Rui-Zhen Huang, Hong-Hao Song, Limei Xu, Chengxiang Ding, and Long Zhang.
	Conformal Boundary Conditions of Symmetry-Enriched Quantum Critical Spin Chains.
\href{https://journals.aps.org/prl/abstract/10.1103/PhysRevLett.129.210601}{\newblock {\em Phys. Rev. Lett.} \textbf{129},  210601 (2022).}

\bibitem{PhysRevA.109.062226}
Hao-Long Zhang1, Han-Ze Li, Sheng Yang, and Xue-Jia Yu.
	Quantum phase transition and critical behavior between the gapless topological phases.
\href{https://journals.aps.org/pra/abstract/10.1103/PhysRevA.109.062226}{\newblock {\em Phys. Rev. A.} {\bf 109}, 062226 (2024).}

\bibitem{PhysRevB.110.045119}
Xue-Jia Yu and Wei-Lin Li.
	Fidelity susceptibility at the Lifshitz transition between the noninteracting topologically distinct quantum critical points.
\href{https://journals.aps.org/prb/abstract/10.1103/PhysRevB.110.045119}{\newblock {\em Phys. Rev. B.} {\bf 110}, 045119 (2024).}

\bibitem{yu2024universal}
Yu, Xue-Jia and Yang, Sheng and Lin, Hai-Qing and Jian, Shao-Kai.
	Universal entanglement spectrum in gapless symmetry protected topological states.
\href{https://journals.aps.org/prl/abstract/10.1103/PhysRevLett.133.026601}{\newblock {\em Phys. Rev. Lett.} {\bf 133}, 026601 (2024).}

\bibitem{PhysRevA.110.022212}
Wen-Hao Zhong, Wei-Lin Li, Yong-Chang Chen, and Xue-Jia Yu.
	Topological edge modes and phase transitions in a critical fermionic chain with long-range interactions.
\href{https://journals.aps.org/pra/abstract/10.1103/PhysRevA.110.022212}{\newblock {\em Phys. Rev. A.} {\bf 110}, 022212 (2024).}

\bibitem{zhou2025gapless}
Sheng Yang, Hai-Qing Lin and Xue-Jia Yu.
	Gapless topological behaviors in a long-range quantum spin chain.
\href{https://www.nature.com/articles/s42005-025-01947-z}{\newblock {\em Communications Physics} {\bf 8}, 27 (2025).}

\bibitem{xue2025gapless}
Xue-Jia Yu, Sheng Yang, Shuo Liu, Hai-Qing Lin, Shao-Kai Jian.
	Gapless Symmetry-Protected Topological States in Measurement-Only Circuits.
\href{https://arxiv.org/abs/2501.03851}{\newblock {\em 	arXiv:2501.03851 [cond-mat.str-el]} (2025).}

\bibitem{xue2025exploring}
Xue-Jia Yu, Sheng Yang, Shuo Liu, Hai-Qing Lin, Shao-Kai Jian.
	Exploring nontrivial topology at quantum criticality in a superconducting processor.
\href{https://arxiv.org/abs/2501.04679}{\newblock {\em 		arXiv:2501.04679 [quant-ph]} (2025).}

\bibitem{zhou2025topological}
Longwen Zhou, Jiangbin Gong and Xue-Jia Yu.
	Topological edge states at Floquet quantum criticality.
\href{https://www.nature.com/articles/s42005-025-02137-7}{\newblock {\em Communications Physics} {\bf 8}, 214 (2025).}

\bibitem{yang2025deconfined}
Sheng Yang, Fu Xu, Da-Chuan Lu, Yi-Zhuang You, Hai-Qing Lin and Xue-Jia Yu.
	Deconfined criticality as intrinsically gapless topological state in one dimension.
\href{https://arxiv.org/abs/2503.01198}{\newblock {\em arXiv:2503.01198 [cond-mat.str-el]} (2025).}

\bibitem{kumar2025topological}
Ranjith R Kumar and Hideaki Obuse.
	Topological transition between gapless phases in quantum walks.
\href{https://arxiv.org/abs/2504.05023}{\newblock {\em arXiv:2504.05023 [quant-ph]} (2025).}

\bibitem{cv5q-8t25}
Wen-Hao Zhong, Hai-Qing Lin, and Xue-Jia Yu
	\newblock Quantum entanglement of fermionic symmetry-enriched quantum critical points in one dimension.
\href{https://journals.aps.org/prb/abstract/10.1103/cv5q-8t25}{\newblock {\em Phys. Rev. B.} \textbf{112}, 075129 (2025).}

\bibitem{haldane1988model}
F. D. M. Haldane.
	\newblock Model for a quantum Hall effect without Landau levels: Condensed-matter realization of the" parity anomaly".
\href{https://journals.aps.org/prl/abstract/10.1103/PhysRevLett.61.2015}{\newblock {\em Phys. Rev. Lett.} \textbf{61}, 2015 (1988).}

\bibitem{kane2005quantum}
Charles L Kane and Eugene J Mele.
	\newblock Quantum spin Hall effect in graphene.
\href{https://journals.aps.org/prl/abstract/10.1103/PhysRevLett.95.226801}{\newblock {\em Phys. Rev. Lett.} \textbf{95}, 226801 (2005).}

\bibitem{wang2017topological}
Jing Wang and Shou-Cheng Zhang.
	\newblock Topological states of condensed matter.
\href{https://www.nature.com/articles/nmat5012}{\newblock {\em Nat. Mater.} \textbf{16}, 1062--1067 (2017).}

\bibitem{thouless1982quantized}
D. J. Thouless, M. Kohmoto, M. P. Nightingale, and M. den Nijs.
	\newblock Quantized Hall conductance in a two-dimensional periodic potential.
\href{https://journals.aps.org/prl/abstract/10.1103/PhysRevLett.49.405}{\newblock {\em Phys. Rev. Lett.} \textbf{49}, 405 (1982).}

\bibitem{hatsugai_chern_1993}
Y. Hatsugai. {Chern number and edge states in the integer quantum Hall effect}.
  \href{http://dx.doi.org/10.1103/PhysRevLett.71.3697}{{\em Phys. Rev. Lett.} {\bf 71}, 3697 (1993)}.

\bibitem{ryu_topological_2002}
S. Ryu and Y. Hatsugai. {Topological origin of zero-energy edge states in particle-hole symmetric systems}.
  \href{http://dx.doi.org/10.1103/PhysRevLett.89.077002}{{\em Phys. Rev. Lett.} {\bf 89}, 077002 (2002)}.

\bibitem{teo_topological_2010}
J. Teo and C. Kane. {Topological defects and gapless modes in insulators and superconductors}. 
\href{http://dx.doi.org/10.1103/PhysRevB.82.115120}{{\em Phys. Rev. B} {\bf 82}, 115120 (2010)}.

\bibitem{hasan2010colloquium}
M~Zahid Hasan and Charles~L Kane.
	\newblock Colloquium: topological insulators.
\href{https://journals.aps.org/rmp/abstract/10.1103/RevModPhys.82.3045}{\newblock {\em Rev. Mod. Phys.} \textbf{82}, 3045 (2010).}

\bibitem{continentino2020finite}
Mucio~A Continentino, Sabrina Rufo, and Griffith~M Rufo.
\newblock Finite size effects in topological quantum phase transitions.
\href{https://link.springer.com/chapter/10.1007/978-3-030-35473-2_12}{\newblock {\em Strongly Coupled Field Theories for Condensed Matter and Quantum Information Theory}}, Springer Proceedings in Physics 239 (2020).

\bibitem{rufo2019multicritical}
Rufo, S., Lopes, N., Continentino, M.A. and Griffith, M.A.R.
	\newblock Multicritical behavior in topological phase transitions.
\href{https://journals.aps.org/prb/abstract/10.1103/PhysRevB.100.195432}{\newblock {\em Phys. Rev. B.} \textbf{100}, 195432 (2019).}

\bibitem{10.21468/SciPostPhys.3.3.021}
Ville Lahtinen, Jiannis K. Pachos.
\newblock A Short Introduction to Topological Quantum Computation.
\href{https://scipost.org/10.21468/SciPostPhys.3.3.021}{\newblock {\em SciPost Phys. } \textbf{3}, 021 (2017).}

\bibitem{nayak_non-abelian_2008}
C. Nayak, S. Simon, A. Stern, M. Freedman, and S. Das~Sarma. {Non-Abelian anyons and topological quantum computation}.
  \href{http://dx.doi.org/10.1103/RevModPhys.80.1083}{{\em Rev. Mod. Phys.} {\bf 80},
  1083 (2008)}.

\bibitem{stern_topological_2013}
A. Stern and N. Lindner. {Topological quantum computation -- From basic concepts to first experiments}.
  \href{http://dx.doi.org/10.1126/science.1231473}{{\em Science} {\bf 339}, 1179
  (2013)}.

\bibitem{das-sarma_majorana_2015}
S. Das~Sarma, M. Freedman, and C. Nayak. {Majorana zero modes and topological quantum computation}. 
\href{http://dx.doi.org/10.1038/npjqi.2015.1}{{\em npj
  Quantum Inf.} {\bf 1}, 15001 (2015)}.

\bibitem{roy_topological_2017}
A. {Roy} and D.~P. {DiVincenzo}. {Topological quantum computing}.
  \href{https://arxiv.org/abs/1701.05052}{{\em arXiv:1701.05052 [quant-ph]}} (2017).

\bibitem{lahtinen_a-short_2017}
V. Lahtinen and J. Pachos. {A short introduction to topological quantum computation}. 
\href{http://dx.doi.org/10.21468/SciPostPhys.3.3.021}{{\em SciPost Phys.} {\bf 3}, 021 (2017)}.

\bibitem{stanescu_introduction_2017}
T.~D. Stanescu, \href{https://www.taylorfrancis.com/books/mono/10.1201/9781003226048/introduction-topological-quantum-matter-quantum-computation-tudor-stanescu}{Introduction to topological quantum matter \& quantum
  computation} (Taylor \& Francis, 2017).

\bibitem{beenakker_search_2020}
C.~W.~J. Beenakker. {Search for non-Abelian Majorana braiding statistics in superconductors}.
\href{http://dx.doi.org/10.21468/SciPostPhysLectNotes.15}{{\em SciPost Phys. Lect. Notes} \textbf{15} (2020)}.

\bibitem{marra_Majorana_2022}
P. Marra. {Majorana nanowires for topological quantum computation}.
  \href{http://dx.doi.org/10.1063/5.0102999}{{\em Journal of Applied Physics} {\bf
  132}, 231101 (2022)}.

\bibitem{su_solitons_1979}
W.~P. Su, J.~R. Schrieffer, and A.~J. Heeger. {Solitons in Polyacetylene}.
  \href{http://dx.doi.org/10.1103/PhysRevLett.42.1698}{{\em Phys. Rev. Lett.} {\bf
  42}, 1698 (1979)}.

\bibitem{kitaev_unpaired_2001}
A.~Y. Kitaev. {Unpaired Majorana fermions in quantum wires}.
  \href{http://dx.doi.org/10.1070/1063-7869/44/10s/s29}{{\em Phys. Usp.} {\bf 44},
  131 (2001)}.

\bibitem{niu_majorana_2012}
Y. Niu, S. Chung, C.-H. Hsu, I. Mandal, S. Raghu, and S. Chakravarty. {Majorana zero modes in a quantum Ising chain with longer-ranged interactions},
  \href{http://dx.doi.org/10.1103/PhysRevB.85.035110}{{\em Phys. Rev. B} {\bf 85},
  035110 (2012)}.

\bibitem{hsu2020topological}
	Hsiu-Chuan Hsu and Tsung-Wei Chen.
	\newblock Topological anderson insulating phases in the long-range su-schrieffer-heeger model.
\href{https://journals.aps.org/prb/abstract/10.1103/PhysRevB.102.205425}{\newblock {\em Phys. Rev. B.} \textbf{102}, 205425 (2020).}

\bibitem{vodola_kitaev_2014}
D. Vodola, L. Lepori, E. Ercolessi, A. Gorshkov, and G. Pupillo. {Kitaev chains with long-range pairing}.
  \href{http://dx.doi.org/10.1103/PhysRevLett.113.156402}{{\em Phys. Rev. Lett.} {\bf
  113}, 156402 (2014)}.

\bibitem{viyuela_topological_2016}
O. Viyuela, D. Vodola, G. Pupillo, and M.~A. Martin-Delgado. {Topological massive Dirac edge modes and long-range superconducting Hamiltonians}.
  \href{http://dx.doi.org/10.1103/PhysRevB.94.125121}{{\em Phys. Rev. B} {\bf 94},
  125121 (2016)}.

\bibitem{alecce_extended_2017}
A. Alecce and L. Dell'Anna. {Extended Kitaev chain with longer-range hopping and pairing}. 
\href{http://dx.doi.org/10.1103/PhysRevB.95.195160}{{\em Phys. Rev. B} {\bf 95}, 195160 (2017)}.


\bibitem{chen2019universality}
	Wei Chen and Andreas~P Schnyder.
	\newblock Universality classes of topological phase transitions with higher-order band crossing.
	\href{https://iopscience.iop.org/article/10.1088/1367-2630/ab2a2d}{\newblock {\em New J. Phys.} \textbf{21}, 073003 (2019).}	

\bibitem{PhysRevB.101.035109}
Ziyu Tao, Tongxing Yan, Weiyang Liu, Jingjing Niu, Yuxuan Zhou, Libo Zhang, Hao
Jia, Weiqiang Chen, Song Liu, Yuanzhen Chen, and Dapeng Yu.
	\newblock Simulation of a topological phase transition in a kitaev chain with long-range coupling using a superconducting circuit.
\href{https://journals.aps.org/prb/abstract/10.1103/PhysRevB.101.035109}{\newblock {\em Phys. Rev. B.} \textbf{101}, 035109, (2020).}

\bibitem{niu2021simulation}
Jingjing Niu, Tongxing Yan, Yuxuan Zhou, Ziyu Tao, Xiaole Li, Weiyang Liu, Libo
Zhang, Hao Jia, Song Liu, Zhongbo Yan, et~al.
\newblock Simulation of higher-order topological phases and related topological phase transitions in a superconducting qubit.
\href{https://www.sciencedirect.com/science/article/pii/S2095927321001766}{\newblock {\em Sci. Bull.} \textbf{66}, 1168--1175 (2021).}

\bibitem{goldman2016topological}
Nathan Goldman, Jan~C Budich, and Peter Zoller.
\newblock Topological quantum matter with ultracold gases in optical lattices.
\href{https://www.nature.com/articles/nphys3803}{\newblock {\em Nat. Phys.} \textbf{12}, 639--645 (2016).}

\bibitem{meier2016observation}
Eric~J Meier, Fangzhao~Alex An, and Bryce Gadway.
\newblock Observation of the topological soliton state in the su--schrieffer--heeger model.
\href{https://www.nature.com/articles/ncomms13986}{\newblock {\em Nat. Commun.} \textbf{7}, 1--6 (2016).}

%\bibitem{PhysRevA.82.033429}
%Takuya Kitagawa, Mark S. Rudner, Erez Berg, and Eugene Demler.
%\newblock Exploring topological phases with quantum walks.
%\href{https://journals.aps.org/pra/abstract/10.1103/PhysRevA.82.033429}{\newblock {\em Phys. Rev. A.} \textbf{82}, 033429 (2010).}

\bibitem{PhysRevB.88.121406}
Janos K. Asboth and Hideaki Obuse.
\newblock Bulk-boundary correspondence for chiral symmetric quantum walks.
\href{https://journals.aps.org/prb/abstract/10.1103/PhysRevB.88.121406}{\newblock {\em Phys. Rev. B} \textbf{88}, 121406(R) (2013).}

\bibitem{PhysRevLett.118.130501}
V.V. Ramasesh, E. Flurin, M. Rudner, I Siddiqi, and N. Y. Yao.
\newblock Direct Probe of Topological Invariants Using Bloch Oscillating Quantum Walks.
\href{https://journals.aps.org/prl/abstract/10.1103/PhysRevLett.118.130501}{\newblock {\em Phys. Rev. Lett.} \textbf{118}, 130501 (2017).}

\bibitem{lu2022exact}
Zhanpeng Lu, Zhihao Xu, Yunbo Zhang.
\newblock Exact Mobility Edges and Topological Anderson Insulating Phase in a Slowly Varying Quasiperiodic Model.
\href{https://onlinelibrary.wiley.com/doi/10.1002/andp.202200203}{\newblock {\em Annalen der Physik} \textbf{534}, 2200203 (2022).}

\bibitem{SIRCAR2026131192}
Sayan Sircar.
\newblock Disorder driven topological phase transitions in 1D mechanical quasicrystals.
\href{https://www.sciencedirect.com/science/article/abs/pii/S0375960125009727?via%3Dihub}{\newblock {\em Phys. Lett. A} \textbf{567}, 131192 (2026).}
\end{thebibliography}
\end{document}